\let\PyraimdKernelLabel\label
\documentclass[aps,10pt,twocolumn,amsmath,amssymb,superscriptaddress,prx,floatfix]{revtex4-2}
\usepackage{graphicx}
\usepackage{bm}
\usepackage{booktabs}
\usepackage{xcolor}
\usepackage{siunitx}
\let\label\PyraimdKernelLabel
\makeatletter\let\ltx@label\label\makeatother
\usepackage[colorlinks=true,linkcolor=blue!50!black,citecolor=blue!50!black,urlcolor=blue!50!black]{hyperref}
\hypersetup{pdftitle={The energetics of force errors in machine-learned molecular dynamics}}
\makeatletter
\newcommand{\statementheading}[1]{%
  \@startsection{subsection}{2}{\z@}{12pt}{4pt}%
    {\normalfont\small\bfseries\centering}*{#1}}
\makeatother
\newcommand{\eps}{\varepsilon_{\mathrm{acc}}}
\newcommand{\E}{\mathbb{E}}
\newcommand{\Prob}{\mathbb{P}}

\begin{document}
\raggedbottom
\title{The energetics of force errors in machine-learned molecular dynamics}
\author{Peng Kang}
\affiliation{School of Materials Science and Engineering, Beihang University, Beijing 100191, China}
\affiliation{National Key Laboratory of Artificial Intelligence for Material Science, Beihang University, Beijing 100191, China}
\affiliation{Center for the Physics of Materials and Department of Physics, McGill University, Montreal, Quebec H3A 2T8, Canada}
\affiliation{Nanoacademic Technologies Inc., Suite 802, 666 Sherbrooke West, Montreal, Quebec H3A 1E7, Canada}
\author{Da Wan}
\affiliation{School of Materials Science and Engineering, Beihang University, Beijing 100191, China}
\affiliation{National Key Laboratory of Artificial Intelligence for Material Science, Beihang University, Beijing 100191, China}
\author{Shulin Bai}
\affiliation{School of Materials Science and Engineering, Beihang University, Beijing 100191, China}
\affiliation{National Key Laboratory of Artificial Intelligence for Material Science, Beihang University, Beijing 100191, China}
\author{Vincent Michaud-Rioux}
\affiliation{Nanoacademic Technologies Inc., Suite 802, 666 Sherbrooke West, Montreal, Quebec H3A 1E7, Canada}
\author{Zhen Li}
\affiliation{School of Materials Science and Engineering, Beihang University, Beijing 100191, China}
\affiliation{National Key Laboratory of Artificial Intelligence for Material Science, Beihang University, Beijing 100191, China}
\author{Yu Liu}
\affiliation{School of Materials Science and Engineering, Beihang University, Beijing 100191, China}
\affiliation{National Key Laboratory of Artificial Intelligence for Material Science, Beihang University, Beijing 100191, China}
\author{Lei Zheng}
\email{zhenglei@buaa.edu.cn}
\affiliation{School of Materials Science and Engineering, Beihang University, Beijing 100191, China}
\affiliation{National Key Laboratory of Artificial Intelligence for Material Science, Beihang University, Beijing 100191, China}
\author{Li-Dong Zhao}
\email{zhaolidong@buaa.edu.cn}
\affiliation{School of Materials Science and Engineering, Beihang University, Beijing 100191, China}
\affiliation{National Key Laboratory of Artificial Intelligence for Material Science, Beihang University, Beijing 100191, China}
\date{September 8, 2026}
\begin{abstract}
The energetic effect of a force error depends on atomic motion. We establish a directional residual-work coefficient combining directional curvature mismatch with the spatial distribution of the residual response. For conservative potentials force-matched at an anchor, it determines the leading signed work at the first crossing of a small force-error budget. At a 474-atom lithium--electrolyte interface, predictions fixed before future reference evaluations differ from measurements by less than 4.6\% of predicted work across 24 prescribed endpoints. Changing only the initial velocity direction at fixed structure and initial total kinetic energy reverses the force--work ranking. At the same admitted time of 0.25\,fs, one direction gives an 11.6\% larger maximum force residual but 36.1\% less work. The reversal recurs at a second structure. The framework connects force tolerances to reference-energy transfer, providing a physical basis for potential assessment and adaptive reference allocation.
\end{abstract}
\maketitle

\begin{figure*}[!tp]
\centering
\includegraphics[width=\textwidth]{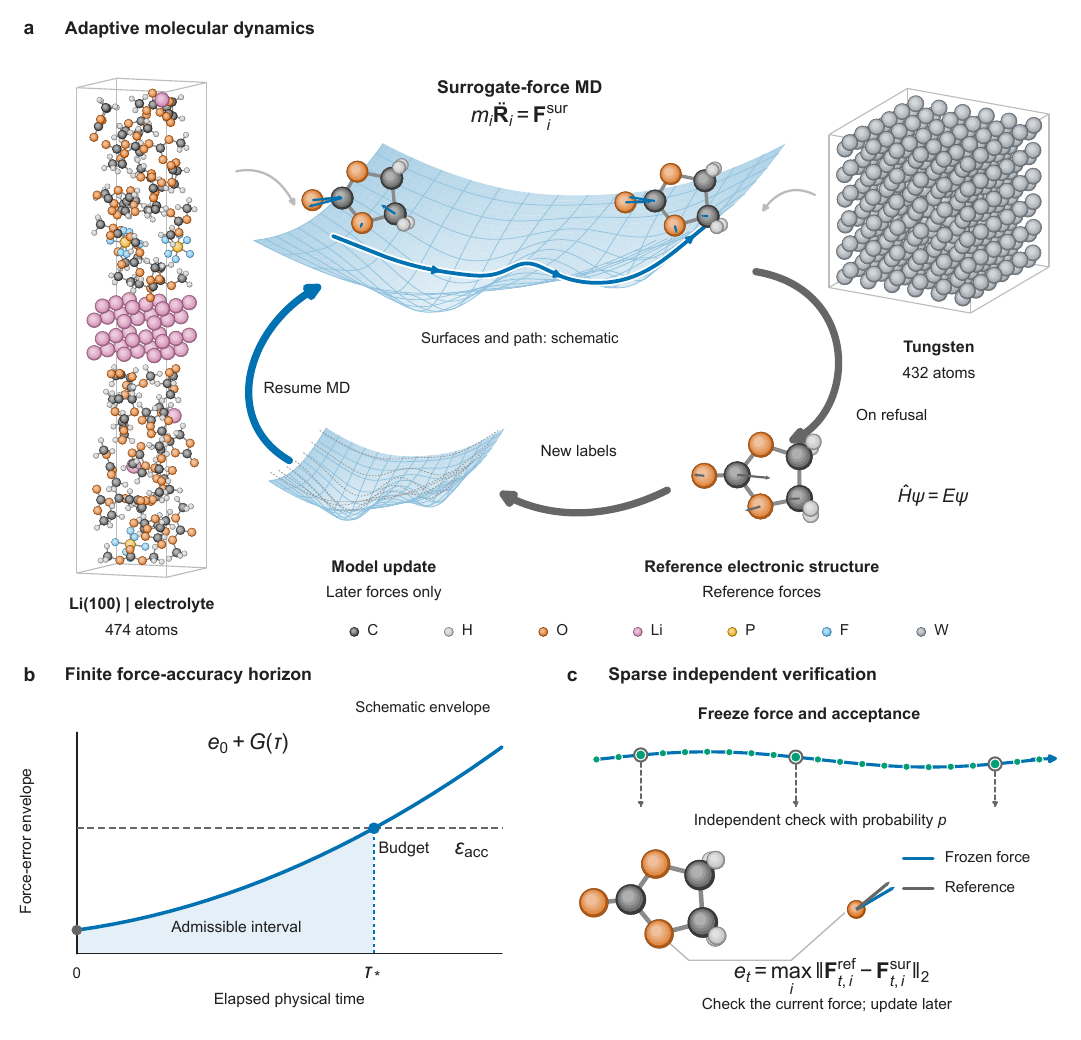}
\caption{\label{fig:materials-principle}\textbf{Reference calculations anchor a finite interval of surrogate-driven atomic motion.} Blue denotes surrogate propagation or its force-error envelope, gray the reference, and green accepted decisions. \textbf{a,} The 474-atom Li(100)--electrolyte interface and a 432-atom tungsten thermal-spike configuration surround the molecular-dynamics cycle. Ethylene carbonate (EC) geometries and surrogate-force arrows use archived interface steps 37 and 42; the tungsten geometry is a committee-sampled configuration at 50 fs. The potential surfaces and connecting path are schematic. \textbf{b,} A schematic residual envelope defines a force-accuracy horizon while it stays below the force budget and within its validity interval, with the model fixed. \textbf{c,} The schematic check stream illustrates independent checks after the force proposal and admission decision have been fixed. An accepted step retains its proposed force; reference labels update later forces. The EC and oxygen-force comparison uses retrospective interface labels at step 128. Atomic images use archived coordinates and rigid projections; force arrows retain their projected source directions with a common linear scale within each illustration.}
\end{figure*}

\section{Introduction}
Understanding how microscopic interactions drive the evolution of matter is a central problem in physics, chemistry and materials science. Atomic rearrangements underlie diffusion, chemical reactions and structural transformations across a wide range of length and time scales. Molecular dynamics provides a microscopic route to these phenomena by following atomic trajectories generated by interatomic forces. Its predictive reach depends on describing these interactions accurately while accessing the length and time scales required by the process of interest. Ab initio molecular dynamics (AIMD) evaluates forces from the electronic structure as the atomic configuration evolves~\cite{car1985unified}. Repeated electronic-structure calculations, however, constrain the system sizes and timescales that can be reached. Machine-learned interatomic potentials approximate reference energies and forces at much lower computational cost~\cite{behler2007generalized,bartok2010gaussian,zhang2018deep}. Equivariant architectures improve the accuracy and efficiency of these models~\cite{batatia2022mace,bochkarev2024grace,li2026dpa4}, while pretrained potentials extend their coverage across material chemistries~\cite{chen2022universal,deng2023chgnet,batatia2022foundation}. Large reference datasets support materials discovery and increasingly diverse molecular, crystalline and surface simulations~\cite{merchant2023gnome,barros2026omat24,mazitov2025petmad,kim2026sevennet}. Extending the reach of these simulations requires understanding how approximations in the forces affect the physics along a trajectory.

Force accuracy is routinely used to assess a potential and decide when a simulation needs a new reference calculation. On-the-fly learning uses selected reference evaluations to update the potential~\cite{csanyi2004learn,podryabinkin2017,jinnouchi2019fly,vandermause2020fly,zhang2019dpgen}; uncertainty estimates guide exploration and force acceptance~\cite{kulichenko2023udd,musil2019fast,imbalzano2021uncertainty,hu2022conformal,ho2025conformal}. Benchmarks examine crystal stability, phonons and trajectory observables, linking model assessment to the quantities simulations seek to predict~\cite{riebesell2025matbench,loew2025phonons,fu2023forces}. Earlier studies distinguished force accuracy from energy conservation and connected conservative force construction and potential smoothness to reliable dynamics~\cite{li2015fly,bigi2025dark,fu2025smooth}. These advances motivate a further question: what determines the energy transferred by a residual force as motion carries the system away from a reference configuration? The residual is the difference between surrogate and reference forces. Its magnitude measures a local discrepancy, while its projection onto atomic motion determines the work that changes the reference energy. A useful force tolerance must therefore be interpreted together with the motion through which that error acts.

Here we derive a quantitative relation between a maximum-atom force-error budget and this signed residual work. For conservative potentials matched in force at a reference configuration, we define the directional residual-work coefficient from the local curvature mismatch. It combines the signed curvature along motion with the spatial distribution of the residual response. At the first crossing of a small force-error budget, the coefficient determines the leading work through a quadratic budget dependence. This relation separates the scale set by the force tolerance from the directional response of the reference and surrogate potentials. Conservation of the surrogate Hamiltonian therefore permits a direction-dependent change in reference energy. The definition uses the reference--surrogate pair and atomic motion, independently of the potential architecture. A bound on the growth of this residual response supplies a force-accuracy horizon for propagation (Fig.~\ref{fig:materials-principle}), while independent reference checks track accepted force errors as the model evolves~\cite{xu2024active,howard2021confidence}.

A 474-atom lithium--electrolyte interface demonstrates the physical consequence of this directional dependence. At fixed structure and initial total kinetic energy, changing only the initial velocity direction reverses the force--work ranking. At the same admitted time of 0.25\,fs, one direction gives an 11.6\% larger maximum force residual but 36.1\% less work. The reversal recurs at a second structure. Predictions fixed before future reference evaluations differ from the measured work by less than 4.6\% of predicted work across 24 prescribed early-time endpoints. Local directional force probes estimate the coefficient without constructing a full Hessian. The resulting connection between force tolerances, atomic motion and reference-energy transfer gives potential assessment and adaptive reference allocation a common physical basis.

\section{Residual work and force-accuracy horizons}
\label{sec:physical-principle}
We first connect the force-error budget to residual work and to the interval available for surrogate propagation. The resulting relations guide the controlled tests of reference allocation and the molecular and material calculations that follow.

\subsection{Directional residual work}
For atomic positions $X$, a proposed surrogate force $\bm F^{\rm sur}$ is compared with a specified reference force $\bm F^{\rm ref}$. We use the largest atomic force residual,
\begin{equation}
 e(X)=\max_i\|\bm F_i^{\rm sur}(X)-\bm F_i^{\rm ref}(X)\|_2,
 \label{eq:main-force}
\end{equation}
and require $e\le\eps$ during force-controlled propagation. This maximum-atom norm resolves the largest local discrepancy and retains its dependence on atom count.

During continuous Newtonian propagation under surrogate forces, the change of the reference Hamiltonian is the work performed by the residual:
\begin{equation}
 \frac{dH_{\rm ref}}{dt}
 =\sum_i\bm v_i\cdot(\bm F_i^{\rm sur}-\bm F_i^{\rm ref}).
 \label{eq:main-work}
\end{equation}
The residual magnitude controls force accuracy, while its projection onto motion controls energy transfer. To isolate the effect of curvature, we add a constant reference-force correction to a fixed base potential. This matches the force exactly at the anchor. The work along the subsequent segment equals the Taylor remainder of the reference-minus-base potential, so curvature mismatch determines its leading sign and magnitude.

Let $H_\Delta$ be the Hessian of $U_{\rm ref}-U_{\rm sur}$ at the force-matched anchor and $\bm u$ the full-configuration Euclidean unit direction of the initial velocity. For $H_\Delta\bm u\ne0$, define the \emph{directional residual-work coefficient},
\begin{equation}
 \mathcal C_{\rm rw}(\bm u)=
 \frac{\bm u^{\mathsf T}H_\Delta\bm u}
      {\|H_\Delta\bm u\|_{\infty,2}^{\,2}}.
 \label{eq:main-work-coefficient}
\end{equation}
The coefficient combines signed curvature with the spatial distribution of the residual response, using the norm in Eq.~\eqref{eq:main-force}. It has units of length squared per energy.

For locally Lipschitz Hessian and velocity, the first time $t_{\eps}$ at which the residual reaches a small budget satisfies
\begin{equation}
 W(t_{\eps})=\frac{\eps^2}{2}\mathcal C_{\rm rw}(\bm u)
 +O(\eps^3).
 \label{eq:main-budget-work}
\end{equation}
The sign of $\mathcal C_{\rm rw}$ gives the sign of the leading reference-energy change. At fixed direction, its value is independent of the initial speed; speed instead sets the leading time to the budget. Appendix~\ref{sec:work} derives the expansion, the underlying work identity and the separate integration and thermostat terms.

\subsection{The propagation interval}
A reference calculation at $t_0$ gives an initial residual $e_0$. With the model fixed during the next interval, a valid growth envelope $e(t_0+\tau)\le e_0+L\tau$ gives
\begin{equation}
 \tau_* = \min\left\{H_{\rm env},\frac{\eps-e_0}{L}\right\},
 \qquad e_0\le\eps,\quad L>0,
 \label{eq:main-horizon}
\end{equation}
where $H_{\rm env}$ is the envelope's applicability interval. The initial margin and the residual growth rate jointly determine the allowed time. A larger speed or a stronger curvature mismatch shortens the interval. The same Hessian $H_\Delta$ controls residual growth through $\|H_\Delta(X)\dot X\|_{\infty,2}$. After retraining, the residual has changed and the horizon must be recomputed or supplied with a bound on the update. Appendix~\ref{sec:horizon} gives the assumptions, limiting cases, and normalized-committee form.

For a committee spread $s$, we use an empirical rule that combines the residual-calibrated score $B=q(s+\delta)$ with an elapsed-time correction during an acceptance streak. The molecular experiments below test these estimates against independent reference forces.

\begin{figure*}[!tp]
\centering
\includegraphics[width=\textwidth]{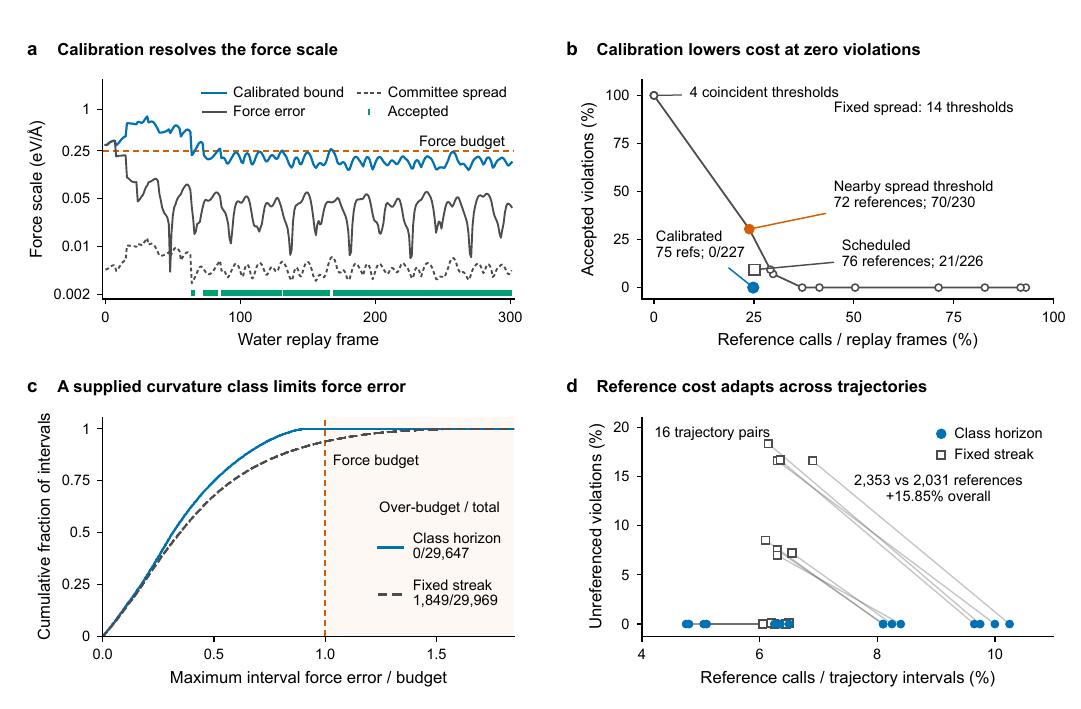}
\caption{\label{fig:materials-validation}\textbf{Residual calibration and force-accuracy horizons reduce force-budget violations.} Blue denotes calibration or the supplied-class horizon and gray the reference comparisons. \textbf{a,} Maximum-atom force error, spread and calibrated bound across all 302 water replay records. The budget is 0.25 eV/\AA; green ticks mark acceptance. The initially infinite bound is omitted. \textbf{b,} Reference-call fraction versus accepted-violation fraction for all 14 measured spread thresholds, calibration and scheduling. The conservative zero-violation threshold uses 112 references; calibration uses 75. Four zero-cost thresholds coincide; one recovered setting has rounded fractions. Lines connect measured settings. \textbf{c,} Empirical cumulative distributions of the exact interval maximum error divided by the 0.05 reduced-unit budget, restricted to unreferenced intervals from 16 held-out oscillator trajectories per policy. The horizon uses the supplied class $|k-1|\le0.2$; held-out $k=1.18$. \textbf{d,} Reference cost and violation fraction for each paired initial condition, with connectors between separate forward trajectories. Fixed $K=18$ was determined from calibration cost. The horizon uses 2353 references against 2031 for the baseline, an increase of 15.85\%. Costs count controller calls; shared development and measurement-only reference evaluations are reported separately in Appendix~\ref{sec:controlled}.}
\end{figure*}

\subsection{An observable record of accepted errors}

After a configuration has been selected for surrogate propagation, an independent coin requests a reference with probability $p$. The measured residual is evaluated on the already fixed force prediction; the label may change later forces and models. If $N$ counts all accepted steps and $D$ counts the over-budget predictions detected by these checks, a prespecified $\lambda>0$ gives the simultaneous upper bound
\begin{equation}
 R\le\min\left\{1,
 \frac{\lambda D+\log(1/\eta)}{N[-\log(1-p+pe^{-\lambda})]}\right\}
 \label{eq:main-verification}
\end{equation}
at all times with $N>0$, with confidence at least $1-\eta$, where $R$ is the actual accepted-violation fraction. The bound allows trajectory dependence and retraining from past checks. The proof appears in Appendix~\ref{sec:verification}.

Before the first detected violation, certifying a target fraction $\alpha$ requires approximately $pN\simeq\log(1/\eta)/\alpha$ checks for small $p$. For example, $p=\eta=0.05$ gives a simultaneous upper count of 58: if 1200 accepted decisions are reached with no detection, the observed record is certified below 4.84\%. A design with a fixed total of 1200 accepted decisions has 60 expected checks, in addition to development and refusal calls (Appendix~\ref{sec:verification}). The growth envelope determines when to propagate, and the independent checks measure how often the accepted forces exceed the budget.

\begin{figure*}[!tp]
\centering
\includegraphics[width=\textwidth]{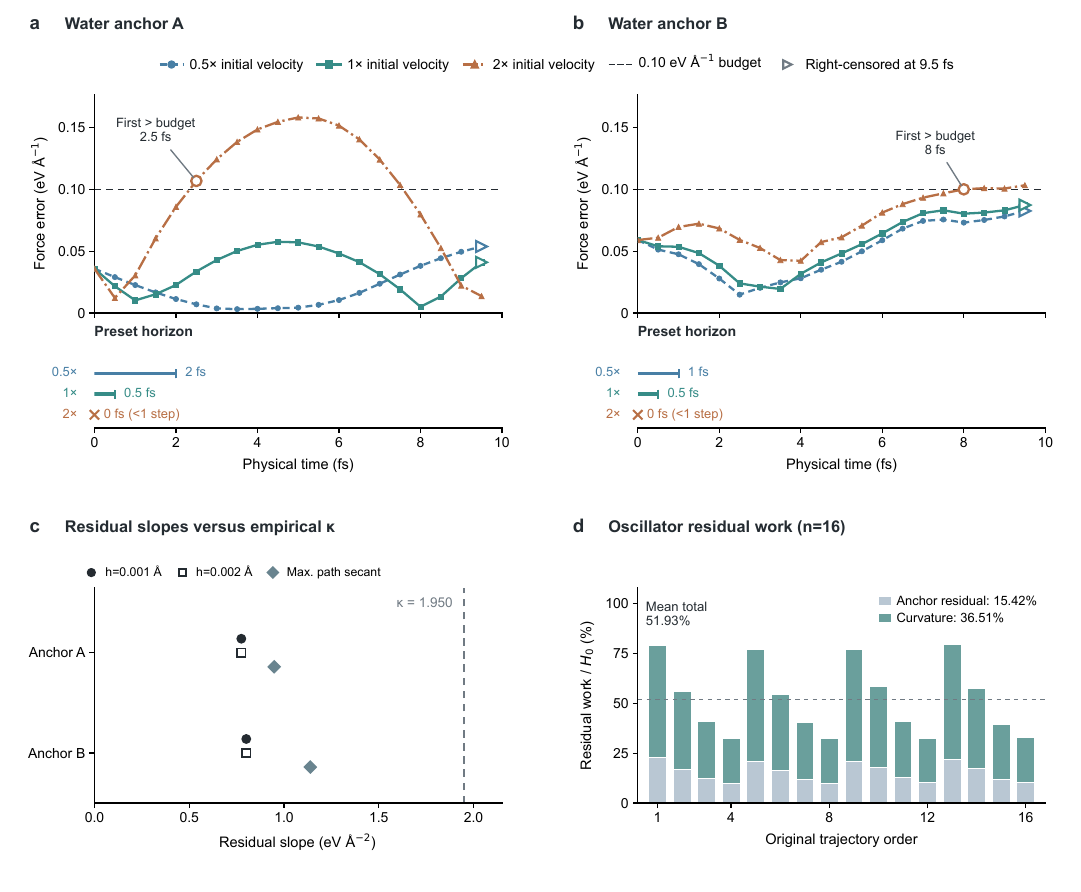}
\caption{\textbf{Velocity dependence of molecular force error and residual work in the oscillator.}
\textbf{a,b,} Maximum-atom Euclidean force error for three velocity factors at each of two water anchors. Lines connect the 20 saved states; the dashed horizontal line is the $0.10\,\mathrm{eV\,\mathring{A}^{-1}}$ budget. Open right-pointing markers identify records with no observed exceedance through 9.5\,fs. Labels mark the first observed exceedances in the $2\times$ interventions. Lower strips give the preset horizons separately from the observed trajectories; 0\,fs means less than one admissible 0.5\,fs forecast step.
\textbf{c,} Norms of the measured central directional residual derivatives at $h=0.001$ and $0.002$\,\AA\ (staggered vertically for visibility), the largest adjacent-state residual-vector secant among the three paths at each anchor, and the recorded empirical $\widehat\kappa$ (dashed). The direction has unit full-configuration Euclidean norm; derivative and secant outputs use the maximum-atom Euclidean norm, giving units of $\mathrm{eV\,\mathring{A}^{-2}}$.
\textbf{d,} Additive residual-work contributions for all 16 held-out class-horizon oscillator trajectories, normalized by each trajectory's initial reference energy $H_0$. The mean anchor-residual and curvature contributions are 15.42\% and 36.51\%, respectively; the dashed line marks their mean sum, 51.93\%. Bars show individual trajectories in original order. The molecular interventions use 122 new reference evaluations; the oscillator bars decompose work along the existing trajectories.}
\label{fig:forward-motion}
\end{figure*}

\section{Force control at limited reference cost}
\label{sec:main-validation}
\subsection{A fully reference-labeled molecular trajectory}
We first compare reference-call policies on 302 configurations from a constant-energy (NVE) trajectory of a single water molecule, initialized at 300 K and driven by a frozen foundation potential. Reference forces are calculated with PBE/def2-SVP at every configuration. A four-member committee processes the configurations chronologically, learning only from labels revealed by its own reference-call policy. The complete reference record gives the actual accepted-error count for each policy.

At $\eps=0.25$ eV/\AA, residual calibration requests 75 reference calculations and accepts 227 configurations with no violations on this stream (Fig.~\ref{fig:materials-validation}). A period-four reference schedule uses 76 calls and has 21 violations among 226 accepted configurations. The nearby measured spread threshold uses 72 calls but has 70 violations among 230 acceptances. A conservative threshold also obtains zero violations, with 112 reference calls.

Each policy selects its own training configurations from the same stream. The comparison therefore tests both which configurations are learned and which forces are accepted. Tighter budgets require more reference calculations, approaching full reference use. Appendix~\ref{app:calibration} gives the additional operating points and calibration details.

\subsection{Atomic motion determines when the reference is needed}
We next test the horizon on trajectories generated by each policy. The unit-mass reference oscillator has force $-kx$ and the surrogate has force $-x+b$, with $b$ updated after a reference measurement. For a declared mismatch $|k-1|\le0.2$, exact propagation of the surrogate gives a speed bound and hence a prospective linear residual envelope. Every policy advances the same physical interval $\Delta t=0.01$; changes in the horizon do not change the integration timestep.

Eight calibration trajectories determine a fixed-streak baseline, $K=18$, before 16 new initial-condition seeds at $k=1.18$ are evaluated. The force-accuracy horizon produces no violations over 29,647 unreferenced intervals; the baseline has 1849/29,969 (6.17\%). It uses 2353 reference calls instead of 2031, an increase of 15.85\% within the prespecified 20\% cost tolerance (Fig.~\ref{fig:materials-validation}).

The residual forces also change the reference energy along these trajectories. The mean final change is $+51.93\%$: curvature mismatch contributes $36.51$ percentage points and the residual left at each anchor contributes $15.42$. Because propagation is exact, these contributions measure residual work without integration error. They show how work accumulates over successive force-controlled intervals. Appendix~\ref{sec:work} gives the derivation; Appendix~\ref{sec:controlled} reports the full controlled study, including the envelope-boundary and statistical tests.

\subsection{Velocity interventions at two molecular anchors}

To test how atomic motion affects force-budget crossings, we scaled the same initial velocity at each of two water configurations by $0.5$, $1$, and $2$, keeping the surrogate fixed. Each record contains 20 force-evaluation states at 0.5\,fs spacing, spanning 0--9.5\,fs; the experiment used 122 new reference evaluations, comprising 114 future trajectory states and eight directional probes. At the $0.10\,\mathrm{eV\,\mathring{A}^{-1}}$ budget, the two $2\times$ trajectories first exceeded the budget at recorded times of 2.5 and 8.0\,fs, respectively, while the other four trajectories had not exceeded the budget by 9.5\,fs (Fig.~\ref{fig:forward-motion}a,b). The preset empirical horizons were $(2.0,0.5,0)$\,fs for anchor A and $(1.0,0.5,0)$\,fs for anchor B, in increasing velocity-factor order. A zero horizon means that the first 0.5\,fs forecast step is not admitted. The forecasts precede the observed crossings and agree with the subsequently generated positions (Appendix~\ref{app:forward-molecular}).

Local directional residual derivatives explain these short horizons. Central differences at the initial anchors, with displacements of 0.001 and 0.002\,\AA\ along the normalized full-configuration velocity, gave residual-derivative vectors differing by 0.00705\% and 0.00969\%. At the smaller displacement, the derivative norms were 0.775 for anchor A and 0.801 for anchor B, in units of $\mathrm{eV\,\mathring{A}^{-2}}$, compared with the recorded empirical value $\widehat\kappa=1.950$ (Fig.~\ref{fig:forward-motion}c). The largest adjacent-state residual-vector secants on the six paths ranged from $0.791$ to $1.139\,\mathrm{eV\,\mathring{A}^{-2}}$. At every recorded point, the residual was below the empirical estimate $e_0+\widehat\kappa\ell$ along the observed polygonal path. Figure~\ref{fig:forward-motion}d gives the complementary oscillator-work decomposition.

\section{Atomic motion connects force accuracy to residual work}
\label{sec:prospective-interface}
\begin{figure*}[!tp]
\centering
\includegraphics[width=0.99\textwidth]{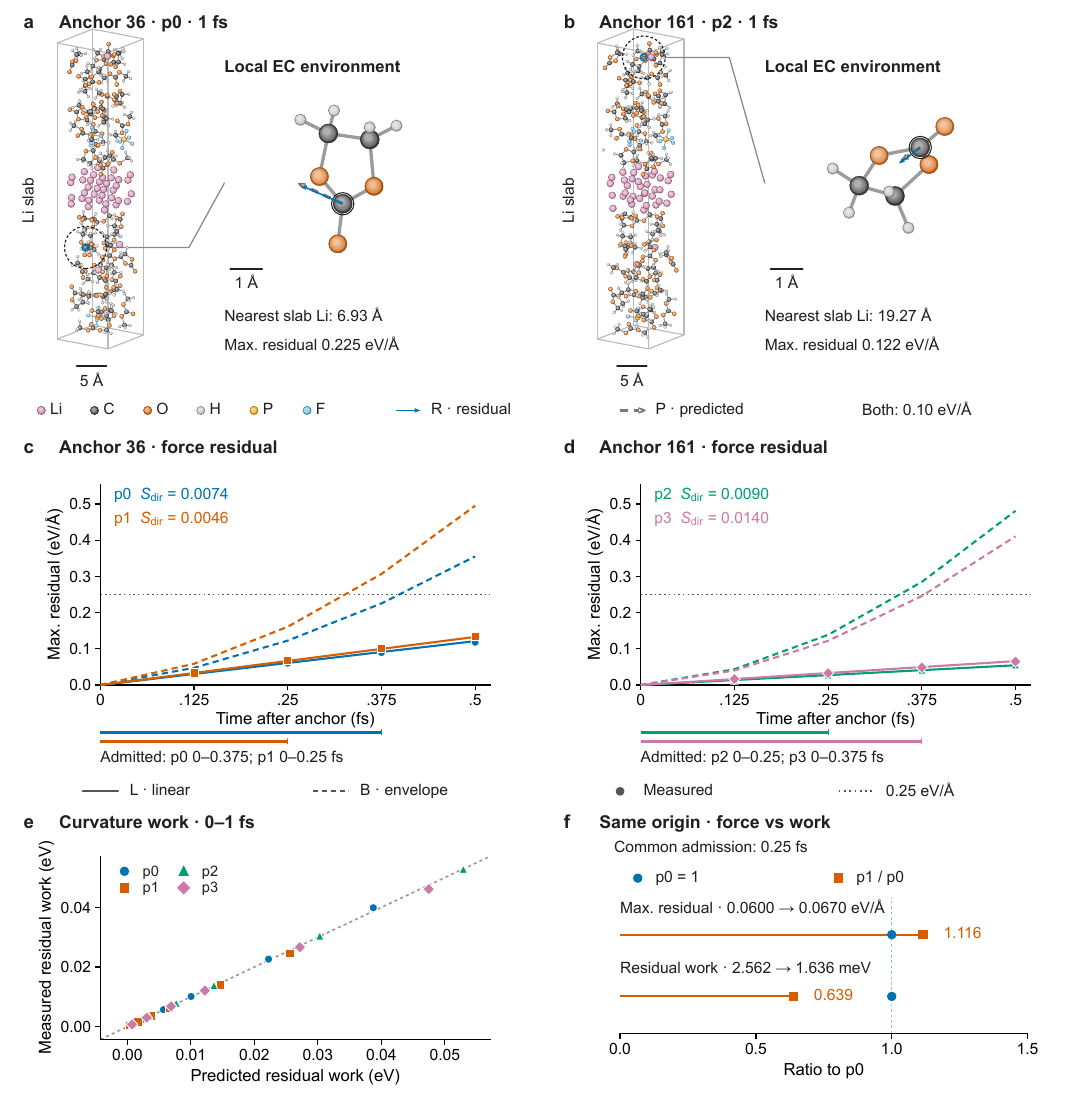}
\caption{\label{fig:prospective-interface}\textbf{Directional force predictions and residual work in a Li(100)--electrolyte interface.}
\textbf{a,b,} Complete cells and local EC environments at the preselected 1-fs endpoints of p0 and p2. Callouts identify the maximum-residual carbons C117 and C157. Blue arrows show $\bm R=\bm F_b+\bm c-\bm F_{\rm ref}$; gray arrows show $\bm P=\alpha\widehat{\bm q}$, with a common projection and force scale. Cell and molecular views each share a length scale.
\textbf{c,d,} Measured residuals compared with frozen linear predictions $L=\|\bm P\|_{\infty,2}$ and empirical scores $B$. Colors identify directions; the dotted line marks 0.25 eV/\AA\ and strips mark the four admission windows (0.375, 0.250, 0.250 and 0.375 fs; ten nonzero reference endpoints). $S_{\rm dir}$ compares squared vector-prediction defects with zero-residual predictions on fixed diagnostic sets (maximum-atom norm).
\textbf{e,} Predicted work $W_{\rm pred}=\tfrac12\alpha^2\bm u\cdot\widehat{\bm q}$ versus measured work at all 24 prescribed endpoints from 0.125 to 1 fs, including points beyond admission. Gray denotes equality.
\textbf{f,} Opposite force--work rankings at 0.25 fs, within both paths' admission windows. Paths p0/p1 have the same initial structure and total kinetic energy but different velocity directions. Orange squares give p1/p0; blue circles mark p0=1. The 1-fs mechanism probes in \textbf{a,b} extend beyond the admission windows.}
\end{figure*}

In the 474-atom Li(100)--electrolyte cell, changing only the initial velocity direction at fixed structure and total kinetic energy reverses the ranking of maximum residual and work (Fig.~\ref{fig:prospective-interface}). At 0.25 fs, both directions at the first origin satisfy their prospective admission rules. The second direction has an 11.6\% larger maximum residual (0.0670 versus 0.0600 eV/\AA) but 36.1\% less work (1.636 versus 2.562 meV). At the second origin, the path with the larger residual likewise has less work: the differences are 19.4\% and 12.7\%, respectively, at a time admitted by both rules. This opposite ordering persists at all six prescribed evaluation times from 0.125 to 1 fs at both origins.

Machine-learning simulations now resolve reaction sequences during solid-electrolyte interphase formation on lithium metal~\cite{takenaka2026}. To examine how force errors develop in such an environment, we use a cell containing a four-layer Li(100) slab, 21 ethylene carbonate (EC) molecules, 17 dimethyl carbonate (DMC) molecules and three LiPF$_6$ units. Two saved structures serve as reference origins, each with two prescribed velocity directions normalized to the total kinetic energy corresponding to 600 K. A constant reference-force correction matches the base potential to the force at each origin. As the atoms move, the residual therefore measures the difference in response of the reference and base force fields. We estimate this response from local directional displacements and use it to predict the residuals and construct empirical growth envelopes. The predictions, evaluation geometries, windows and numerical checks are fixed before future reference calculations, as described in Appendix~\ref{app:interface-prospective}.

At a maximum-atom force budget of 0.25 eV/\AA, the four prospective admission windows are 0.375, 0.250, 0.250 and 0.375 fs. All ten admitted nonzero evaluation times satisfy the budget, with measured residuals of 0.0137--0.0892 eV/\AA. A window is admitted only while both the empirical envelope and its displacement-domain conditions hold throughout the preceding path. On the diagnostic sets fixed separately for the four directions, measured maximum residuals are 0.971--1.026 times their directional predictions, and the squared vector-defect ratios span 0.00457--0.01399. The local derivatives estimate $H_\Delta\bm u$, which determines the leading residual growth and work near the anchor in Eq.~\eqref{eq:main-budget-work}.

We use the same directional response to predict residual work. Figure~\ref{fig:prospective-interface}e compares the predictions with measurements at all 24 prescribed endpoints from 0.125 to 1 fs, including times beyond admission. The measured work differs from its frozen prediction by less than 4.6\% of the predicted value. A separate comparison near the anchor tests the directional residual-work coefficient (Eq.~\eqref{eq:main-work-coefficient}). The two finite-probe estimates $\widehat{\mathcal C}_{\rm rw}$ at the first origin are 1.368 and 0.748 \AA$^2$/eV; the measured ratios $2W/e^2$ at 0.125 fs are 1.406 and 0.750 \AA$^2$/eV. Two preselected 1-fs probes examine the work beyond the admission windows. Their maximum residuals are 0.225 and 0.122 eV/\AA, with residual work values of 39.89 and 52.81 meV at the first and second origins, respectively. The development-stage work predictions are 38.73 and 52.90 meV. Appendix~\ref{app:interface-record} gives the complete prescribed endpoint record through 4 fs.

For the primary 0--1-fs work test at the second origin, direct reference-force integration gives 52.764 meV, agreeing with the endpoint work of 52.809 meV to 0.0850\%. The discrepancy decreases across all three quadrature grids (Appendix~\ref{app:interface-prospective}). Tightening the electronic convergence gives a work sensitivity of 0.0651\% when the change of anchor force is included; halving the dynamics timestep changes the endpoint work by 0.0425\%. These comparisons meet all prescribed numerical-resolution criteria. Residual work accounts for 96.6\% of the reference-Hamiltonian change and is 28.1 times the anchored-Hamiltonian integration drift. Force-field mismatch therefore dominates the energetic change over this segment.

To see which atoms contribute to this work, we integrate each atomic residual force along that atom's actual displacement. At the second-origin 1-fs endpoint, the carbon C157 has the largest residual but contributes $+0.644$ meV, or 1.22\% of the 52.764-meV net force integral, ranking seventeenth among the 474 atomic work contributions. The work is distributed across the electrolyte: the initial EC and DMC atoms contribute 60.0\% and 39.0\% of the same net force integral, respectively. Along the second direction at the first origin, O111 has the largest residual at all four early evaluation times. Its contribution to the 0--0.5-fs integral is $-0.372$ meV, whereas the net residual-force integral is $+6.325$ meV, compared with $+6.343$ meV from endpoint energies. This oxygen has the second-largest absolute atomic work contribution, with the opposite sign to the net work. Thus an atom with a large force error can make a substantial contribution that opposes the total. The sign and magnitude depend on how the residual force projects onto its motion.

\section{Force residuals concentrate in the distorted tungsten spike core}
\label{sec:w-material}
\begin{figure*}[!tp]
\centering
\includegraphics[width=\textwidth]{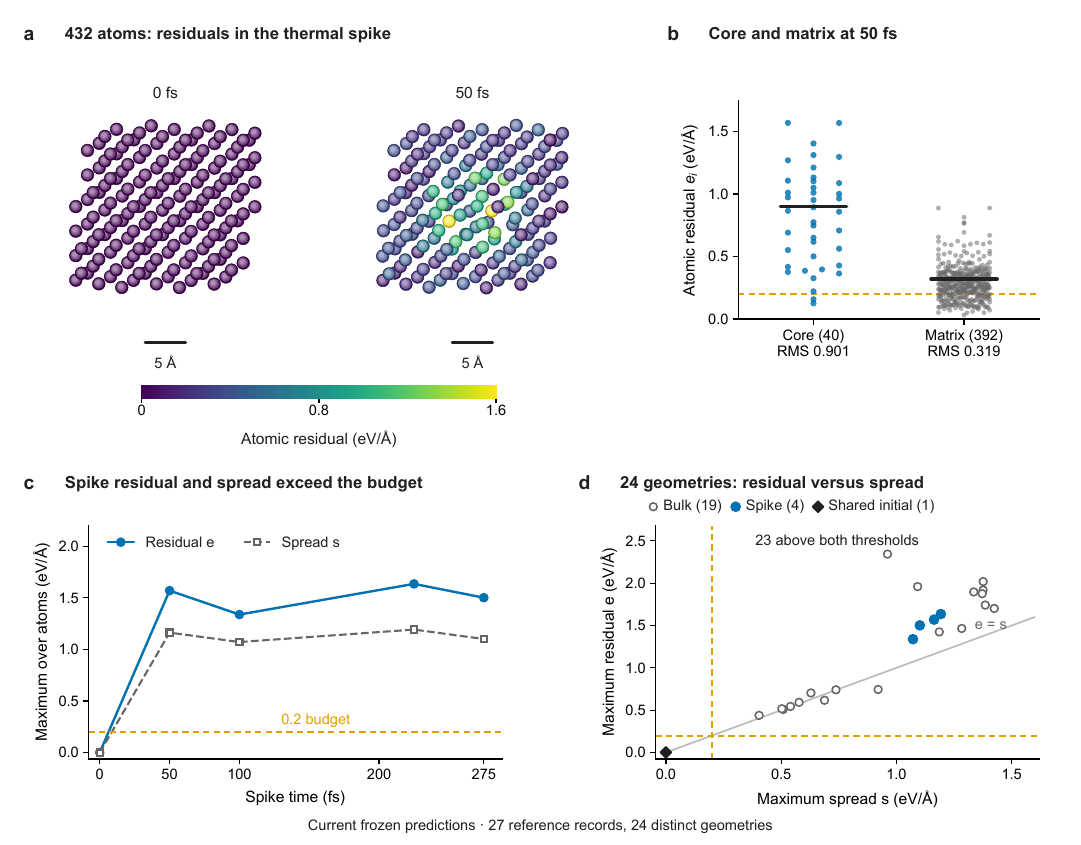}
\caption{\label{fig:materials-tungsten}\textbf{Atomic distortion concentrates the force residual in the tungsten spike core.} Committee predictions are compared with existing Quantum ESPRESSO labels on 24 distinct 432-atom geometries. \textbf{a,} Actual initial and 50-fs spike coordinates, coloured by the atomic residual $e_i=\|\bar{\bm F}_i-\bm F_i^{\rm ref}\|$. A fixed central 6.33-\AA\ display slice contains the same 166 atom identities at both times; projection, magnification and colour scale are shared, with calibrated 5-\AA\ bars. All quantitative panels use complete cells. \textbf{b,} The 40 initial-core and 392 matrix atomic residuals at 50 fs. Each dot is one atom; horizontal offsets separate overlapping dots. Black ticks and labels give regional vector RMS values. \textbf{c,} Maximum residual $e=\max_i e_i$ and maximum spread $s=\max_i\sigma_i$ at the five labeled spike geometries. Lines connect sampled values as visual guides. The shared initial geometry uses its canonical frame-0 reference. \textbf{d,} All 24 distinct geometries: 19 displaced bulk samples, four displaced spike samples and one shared initial lattice. The gray diagonal is $e=s$. Orange dashed lines in \textbf{b--d} mark 0.2 eV/\AA; in \textbf{b} the budget applies to individual atomic residuals. Spread is the empirical population vector RMS $\sigma_i=[\frac12\sum_{k=1}^{2}\|\bm F_{i,k}-\bar{\bm F}_i\|^2]^{1/2}$. Reference labels use $\Gamma$-point PBE-D3 at 70/700 Ry; the committee has no additional D3 correction (Appendix~\ref{app:tungsten-methods}).}
\end{figure*}

Radiation-damage simulations in tungsten require accurate forces across crystalline and highly distorted environments~\cite{byggmastar2019gapw}. Figure~\ref{fig:materials-tungsten} shows that atomic displacements and force residuals concentrate in the initially energized core of a tungsten thermal spike. We use 432-atom body-centered-cubic (bcc) configurations from three bulk-temperature segments and a thermal spike. The spike begins at the same ideal positions as the bulk inputs, with larger momenta assigned to 40 atoms within 5.5 \AA\ of the cell center. The initial perturbation is thus kinetic; atomic motion produces the distortion.

We compare committee forces with 27 successful reference records on 24 distinct geometries; four records correspond to the shared initial lattice. The committee retains its second member's 1\% initialization perturbation. Model weights and the threshold remain fixed; PBE-D3 labels are used only for evaluation (Appendix~\ref{app:tungsten-methods}).

At 50 fs, the initial core had a root-mean-square (RMS) displacement of 0.391 \AA, compared with 0.075 \AA\ in the surrounding matrix. On those same coordinates, the core and matrix force-residual vector RMS values were 0.901 and 0.319 eV/\AA, respectively, and the maximum atomic residual was 1.570 eV/\AA. Regional vector RMS values include all 40 core or 392 matrix atoms (Appendix~\ref{app:tungsten-methods}). The corresponding reference-force vector RMS amplitudes were 6.60 and 1.38 eV/\AA.

To check whether initialization or the D3 correction accounts for the large core residual, we repeat the comparison on the same 50-fs geometry with the original, unperturbed models. Their mean has core and matrix residual RMS values of 0.898 and 0.326 eV/\AA, and maximum residual 1.604 eV/\AA. Removing the reference's additive D3 force, whose maximum is 0.0655 eV/\AA, leaves maximum residual 1.625 eV/\AA\ and core/matrix RMS values of 0.887/0.319 eV/\AA. Neither change removes the large residual in the distorted core. Appendix~\ref{app:tungsten-methods} reports the individual models, D3 checks and shared initial-lattice control.

Across all 24 geometries, the force-residual vector RMS was 0.511 eV/\AA\ and the largest atomic residual was 2.344 eV/\AA. The maximum residual $e$ and maximum committee spread $s$ both exceeded 0.2 eV/\AA\ on every displaced geometry, including all four labeled noninitial spike configurations (Fig.~\ref{fig:materials-tungsten}c,d). The shared ideal-lattice geometry is within both thresholds.

\section{Discussion}
The directional residual-work coefficient quantifies how force errors transfer energy through atomic motion. Together with a residual-growth envelope, it connects the reference-energy change to the interval available for force-controlled propagation.

Force matching removes the initial residual but leaves the curvature difference between the reference and base potentials. Systematic softening studies have identified underestimated potential-energy curvature as a source of error in universal potentials~\cite{deng2025softening}. The signed projection $\bm u^\top H_\Delta\bm u$ determines the leading work, while $\|H_\Delta\bm u\|_{\infty,2}$ determines the leading maximum residual. Their combination in $\mathcal C_{\rm rw}$ sets the leading work at the force boundary (Eq.~\eqref{eq:main-budget-work}). Their different directional dependence explains the force--work reversal at both interface structures. The conservative surrogate forces are energy gradients, and the resolved work substantially exceeds the integration drift.

The atomic work contributions explain why the largest force residual does not determine the total work. The carbon with the largest residual ranks seventeenth in atomic work, which is distributed across the electrolyte. Along another direction, the largest-residual oxygen has the second-largest absolute work contribution, with a sign opposite to the total. Local contributions can therefore reinforce or oppose the net reference-energy change. Under smooth dynamics, a strict force--work reversal at a common time persists for sufficiently small changes of the initial directions, at fixed structure and initial kinetic energy (Appendix~\ref{app:directional-robustness}).

The construction uses reference forces, surrogate forces and atomic motion, and thus applies across materials and potential architectures. For conservative, force-consistent potentials, Eq.~\eqref{eq:main-budget-work} relates the same residual response to force growth and work. Independent randomized checks quantify accepted violations as the dynamics and model adapt~\cite{xu2024active,howard2021confidence}.

\subsection{Scope and outlook}
Longer force-accuracy horizons require information about the residual beyond the initial direction of motion. The oscillator uses a supplied curvature-class bound, while the interface uses empirical local estimates. At 4 fs, the interface paths turn beyond the prescribed transverse domain and their work exceeds the directional predictions by 17.8--70.9\% (Appendix~\ref{app:interface-record}). This behavior points to transverse response and higher-order growth as the next terms to include when extending the prediction interval. Quantitative guarantees require the resulting bounds to remain valid along the generated path and across model updates.

The water tests relate this challenge to reference demand: the empirical horizon improves structural fidelity at the original speed with more reference calls, but does not meet the four-direction screening criterion across tested speeds (Appendix~\ref{app:molecular-feasibility}). Comparisons at matched reference cost would isolate the benefit of work-aware criteria. Further tests can examine reaction rates, transport and ensemble properties against a consistently defined electronic-structure reference. A practical challenge is to estimate residual growth from sparse reference measurements and update the bounds as the atoms leave the local neighborhood. Such bounds would extend propagation intervals at controlled reference cost.

\begin{samepage}
\begin{acknowledgments}
This work was supported by the National Science Fund for Distinguished Young Scholars (No. 51925101), the National Natural Science Foundation of China (Nos. 52450001 and 12104370), the Tianmushan Laboratory Research Project (Nos. TK2024D006 and TK2023C021), and the ``Pioneer'' and ``Leading Goose'' R\&D Program of Zhejiang (Project No. 2024SSYS0084). P. K. would like to thank Compute Canada and the High Performance Computing Center of McGill University for the computer facilities and support at an early stage of this work.
\end{acknowledgments}
\end{samepage}

\appendix
\setcounter{figure}{0}
\renewcommand{\thefigure}{A\arabic{figure}}
\renewcommand{\theHfigure}{appendix.\arabic{figure}}
\setcounter{table}{0}
\renewcommand{\thetable}{A\arabic{table}}
\renewcommand{\theHtable}{appendix.\arabic{table}}
\section{Force errors and accepted violations}
\label{sec:target}
At configuration $X_t$, let $\bm F_t^{\rm ref}$ denote the force from a specified reference calculation and $\bm F_t^{\rm sur}$ the proposed surrogate force, evaluated before any current reference result is revealed. For a system of $n_{\rm at}$ atoms define
\begin{equation}
 e_t=\max_{1\le i\le n_{\rm at}}\|\bm F_{t,i}^{\rm ref}-\bm F_{t,i}^{\rm sur}\|_2,
 \qquad Y_t=\mathbf 1\{e_t>\eps\}.
 \label{eq:error}
\end{equation}
The maximum-atom norm measures the largest atomic force discrepancy. Comparisons between systems of different sizes retain this dependence on atom count.

Let $A_t\in\{0,1\}$ be the admission decision, made before the current verification coin is drawn. The accumulated accepted count, violation count, and realized violation fraction are
\begin{equation}
 N_t=\sum_{j\le t}A_j,\quad V_t=\sum_{j\le t}A_jY_j,
 \quad R_t=V_t/N_t\quad(N_t>0).
 \label{eq:target}
\end{equation}
$V_t$ is generally unobserved because reference forces are expensive. Independent checks provide a probabilistic upper bound on $R_t$, the violation fraction in the accumulated accepted record.

For a committee of $K$ models we use the spread
\begin{equation}
 s_t=\max_i\left[\frac1K\sum_{a=1}^{K}\|\bm F_{t,i}^{(a)}-\overline{\bm F}_{t,i}\|_2^2\right]^{1/2},
 \qquad r_t=\frac{e_t}{s_t+\delta},
 \label{eq:score}
\end{equation}
where $\delta>0$ prevents division by zero. A historical quantile $q_t$ gives the empirical admission score $B_t=q_t(s_t+\delta)$ and the rule $A_t=\mathbf1\{B_t\le\eps\}$.

Even if marginal coverage $\Prob(e_t>B_t)\le\alpha$ holds, for $\Prob(A_t=1)>0$ all it directly implies for admission is
\begin{equation}
 \Prob(e_t>\eps\mid A_t=1)
 \le \min\left\{1,\frac{\alpha}{\Prob(A_t=1)}\right\}.
 \label{eq:selection}
\end{equation}
The conditional bound depends explicitly on the acceptance probability. The original marginal statement also requires exchangeable scores; chronological sampling, reference selection and model updates can alter this condition. Sequential verification uses the independent check coins.

\section{Residual growth between reference calculations}
\label{sec:horizon}
\subsection{An observation limit}
\textbf{Proposition 1 (finite labels do not determine unseen force error).}
Consider a finite set of queried configurations and an accepted configuration $X_*$ distinct from that set, in an open region of configuration space. Within the class of smooth conservative force fields, the queried force labels alone do not impose a finite upper bound on the reference force at $X_*$.

To see this, select a smooth compactly supported function $\psi$ whose support surrounds $X_*$ but excludes all queried configurations and which equals one near $X_*$. Adding
\begin{equation}
 \Delta U(X)=-\bm g\cdot(X-X_*)\psi(X)
 \label{eq:bump}
\end{equation}
to a reference potential leaves every queried force unchanged while changing the force at $X_*$ by any chosen finite vector $\bm g$. The modified force remains conservative and smooth. With the auxiliary randomness fixed, all observed labels remain unchanged. A surrogate-driven loop can therefore make the same decisions and follow the same finite trajectory while its unobserved error at $X_*$ changes. Within this function class, extrapolation from finite labels therefore requires quantitative derivative information.

\subsection{A force-accuracy horizon}
A quantitative regularity assumption changes the conclusion. Freeze the surrogate parameters between reference updates, and suppose the force residual satisfies a known envelope along the actual path,
\begin{equation}
 e(t_0+\tau)\le e_0+G(\tau),\qquad 0\le\tau\le H_{\rm env}.
 \label{eq:envelope}
\end{equation}
Here $e_0$ is a reference-measured upper bound at $t_0$, $G$ is continuous and nondecreasing with $G(0)=0$, and $H_{\rm env}$ is the stated applicability interval. Assume $e_0\le\eps$. The force-accuracy horizon is
\begin{equation}
 \tau_* = \sup\{\tau\in[0,H_{\rm env}]:e_0+G(\tau)\le\eps\}.
 \label{eq:horizon}
\end{equation}
For $G(\tau)=L\tau$, it is $\min\{H_{\rm env},(\eps-e_0)/L\}$ when $L>0$. For $L=0$, the complete applicability interval is covered. If $e_0>\eps$, no acceptance is justified by this envelope, including at $\tau=0$. The expression has units of time; converting it to steps requires the actual integration timestep.

The construction applies to differentiable force models without specifying a potential architecture or a material. Write $\bm R(X)=\bm F^{\rm sur}(X)-\bm F^{\rm ref}(X)$ and let $X(t)$ be an absolutely continuous generated path. A quantitative bound on $\|J_X\bm R(X)\dot X\|_{\infty,2}$ supplies the growth rate. For conservative reference and surrogate forces, $J_X\bm R=H_\Delta$, the Hessian of the reference-minus-surrogate potential. Residual growth thus depends on the curvature mismatch and the atomic velocity. An empirical slope fitted to past errors is a \emph{candidate predictor} of this rate, not a certified upper bound on future motion.

\subsection{Composition across model updates}
\label{app:general-composition}
The same construction accommodates different training procedures and model updates. Let $t_0<t_1<\cdots$ be locally finite update times, with model $j$ fixed on $[t_j,t_{j+1})$; the final upper endpoint is $+\infty$ if updates cease. Assume a fixed continuous reference force, an absolutely continuous configuration path, and $C^1$ residual fields $\bm R_j$ in each path neighborhood, with one-sided endpoint values. Suppose nonnegative integrable bounds satisfy
\begin{align}
 \|J_X\bm R_j(X(s))\dot X(s)\|_{\infty,2}&\le g_j(s),\label{eq:general-growth}\\
 \|\bm F^{\rm sur}_k(X(t_k))-\bm F^{\rm sur}_{k-1}(X(t_k))\|_{\infty,2}&\le d_k.\nonumber
\end{align}
For $j(t)=\max\{j:t_j\le t\}$ and an initial reference-measured upper bound $e_0$, the residual after any update obeys
\begin{align}
 \|\bm R_{j(t)}(X(t))\|_{\infty,2}
 &\le e_0+\sum_{j:t_j<t}\int_{t_j}^{\min(t,t_{j+1})}g_j(s)\,ds\nonumber\\
 &\quad+\sum_{k:t_0<t_k\le t}d_k.\label{eq:general-composition}
\end{align}
This follows by integrating the residual derivative on each fixed-model segment, adding the update jumps, and applying the triangle inequality. The reference force cancels in each jump. The jump norm at a known update configuration can therefore be evaluated from the two surrogate forces without an additional reference calculation. A new reference force can instead reset the anchor and tighten the accumulated bound. The two sums separate residual growth during atomic motion from changes caused by updating model parameters.

For prospective use, each model's growth bound must be available before propagation and valid along its generated path. At an update, both the force jump and the new growth bound enter the admission decision before the new force is applied. An update that raises the bound above the budget is refused, including at the jump endpoint. These conditions apply across atom counts, compositions and architectures; numerical bounds and acceptance rates depend on the implementation. The molecular computations below evaluate the fixed-model specialization with empirical growth estimates.

\subsection{Committee-normalized realization}

For committee normalization, if $r(t_0+\tau)\le r_0+d\tau$ and a nondecreasing upper envelope $\bar s(\tau)$ is also valid throughout the interval, a sufficient condition is
\begin{equation}
 (r_0+d\tau)[\bar s(\tau)+\delta]\le\eps.
 \label{eq:normalized-horizon}
\end{equation}
Both envelopes enter a horizon chosen in advance. When spread is observed at each decision, its measured value can instead be inserted at that time. A fitted slope supplies an empirical estimate to be tested against subsequent references.

\subsection{Calibration during an acceptance streak}
For a fixed, finite $q>0$ during an acceptance streak, define its instantaneous relative margin
\begin{equation}
 m_k=\frac{\eps-q(s_k+\delta)}{q(s_k+\delta)}.
 \label{eq:margin}
\end{equation}
For an accepted step, $m_k\ge0$, and the exact violation condition is
\begin{equation}
 e_k>\eps\ \Longleftrightarrow\ r_k>q(1+m_k).
 \label{eq:exact-leak}
\end{equation}
Under the additional local model $r_k\simeq r_0+d k$ with a constant per-step slope $d$, its approximation is
\begin{equation}
 d k>q m_k+(q-r_0).
 \label{eq:linear-leak}
\end{equation}
The intercept term vanishes when $r_0=q$. Equation~\eqref{eq:exact-leak} diagnoses a crossing using the current spread and reference error; Eq.~\eqref{eq:normalized-horizon} predicts it from an envelope fixed before propagation.

A streak-inflated proposal $B_k=q(s_k+\delta)(1+\gamma k)$ can implement a linear envelope if $r_0\le q$ and $r_k\le r_0+d k$ with $\gamma\ge d/q$. Its validity is conditional on those inequalities. The inflation coefficient $\gamma$ is dimensionless per step and cannot be compared directly to an unnormalized force-error slope.

\section{Sequential verification under adaptive dynamics}
\label{sec:verification}
\subsection{Reference checks after admission}
On an accepted step draw $Z_t\sim\mathrm{Bernoulli}(p)$, independently of all information determining the current state, surrogate force, admission decision, and reference-force error. If $Z_t=1$, evaluate the reference force and record whether the already proposed force violated the budget. Let
\begin{equation}
 D_t=\sum_{j\le t} A_j Z_j Y_j
 \label{eq:detected}
\end{equation}
be the number of detected accepted violations. The verification labels are additional to reference calls on refused steps.

The current surrogate force and admission decision are fixed before the coin and reference result; a checked violation remains an accepted violation in Eqs.~\eqref{eq:target} and \eqref{eq:detected}. The accepted step propagates with its frozen surrogate force. Its reference label may then train the next model, shorten the next horizon or trigger later fallback. For a protocol that replaces the current force after checking, the theorem applies to a separately retained record of pre-check proposals.

\textbf{Proposition 2 (time-uniform upper bound).}
Fix $p\in(0,1)$, $\lambda>0$, and $\eta\in(0,1)$ before examining the verification outcomes. Set
\begin{equation}
 c_\lambda=1-p+p e^{-\lambda},\qquad
 U_t(\lambda,\eta)=\frac{\lambda D_t+\log(1/\eta)}{-\log c_\lambda}.
 \label{eq:bound}
\end{equation}
Under the reference-check independence condition,
\begin{equation}
 \Prob\left\{\forall t:N_t>0,\quad
 R_t\le \min\left(1,\frac{U_t(\lambda,\eta)}{N_t}\right)\right\}\ge1-\eta.
 \label{eq:guarantee}
\end{equation}
The acceptance decisions, dynamics and retraining may otherwise depend arbitrarily on the past, allowing correlated configurations and nonstationary force errors.

The proof is short. Relative to a filtration that includes the state and latent error before each current coin, the process
\begin{equation}
 M_t=\exp(-\lambda D_t)c_\lambda^{-V_t},\qquad M_0=1,
 \label{eq:martingale}
\end{equation}
is a nonnegative martingale. If $A_tY_t=0$, its multiplicative increment is one. Otherwise that increment is $e^{-\lambda Z_t}/c_\lambda$, whose conditional mean is one. Ville's inequality bounds the probability that $M_t$ ever exceeds $1/\eta$ by $\eta$~\cite{howard2021confidence}. Rearranging gives Eq.~\eqref{eq:guarantee} for the accumulated accepted-force record.

The force budget and accepted population are fixed prospectively. The bound remains valid at a data-dependent stopping time. Simultaneous comparisons across budgets, subgroups or a grid of $\lambda$ values can use prespecified error budgets $\eta_j$ with $\sum_j\eta_j\le\eta$; the minimum of the corresponding $\lambda$-bounds then retains the total confidence level.

\subsection{The zero-detection limit and reference cost}
If the verification objective is fixed in advance as certifying a period before the first detected violation, the process
\begin{equation}
 M_t^{(0)}=\mathbf1\{D_t=0\}(1-p)^{-V_t}
 \label{eq:zero-martingale}
\end{equation}
is also a nonnegative martingale. Consequently, with probability at least $1-\eta$, simultaneously at times with $D_t=0$,
\begin{equation}
 V_t<\frac{\log(1/\eta)}{-\log(1-p)}\equiv U_0.
 \label{eq:zero}
\end{equation}
The equivalent sharp event bound is $\Prob(\exists t:D_t=0,\ V_t\ge v)\le(1-p)^v$ for integer $v\ge1$. For a predetermined stream with exactly $v$ violations, the probability of missing them all is exactly $(1-p)^v$. Independent checks therefore give the exponential dependence in Eq.~\eqref{eq:zero}. Once a violation is detected, this particular certificate becomes vacuous; a separately prespecified finite-$\lambda$ certificate remains available with its own error budget. Combining the two requires an allocation of the total error probability.

For example, $p=0.05$ and $\eta=0.05$ give $U_0\simeq58.4$. At 1200 accepted steps with no detected violations, the simultaneous bound is $V_t\le58$, or $R_t\le4.833\ldots\%$. Reaching a fixed target of 1200 accepted decisions requires 60 checks in expectation, taken over all check outcomes. The confidence statement is simultaneous across the accumulated record, with randomness supplied by the checks.

For a fixed prospective target fraction $\alpha$, the approximate zero-detection requirement is
\begin{equation}
 N\gtrsim\frac{\log(1/\eta)}{\alpha[-\log(1-p)]},\qquad
 pN\simeq\frac{\log(1/\eta)}{\alpha}\quad(p\ll1).
 \label{eq:cost}
\end{equation}
Reducing $p$ lengthens the observation horizon at the same leading verification cost. Refusal labels, initialization and retraining add to these checks. At a fixed overall horizon, the expected reference-call count is $\E Q+p\E N$, where $Q$ counts refused steps.

\subsection{Physical interpretation of the count bound}
Each latent accepted violation has the prescribed chance of being observed, including under fully dependent atomic motion and feedback from past labels. The resulting bound describes the accumulated over-budget count for the chosen reference, budget and norm. Energetic effects additionally depend on residual magnitude and direction, as quantified by the work identity below.

\section{Residual work and the reference Hamiltonian}
\label{sec:work}
For continuous dynamics with a time-independent reference potential, define $H_{\rm ref}=\sum_i m_i\|\bm v_i\|^2/2+U_{\rm ref}(X)$. Along a surrogate-driven interval,
\begin{equation}
 \frac{dH_{\rm ref}}{dt}=\sum_i\bm v_i\cdot(\bm F_i^{\rm sur}-\bm F_i^{\rm ref}).
 \label{eq:work-identity}
\end{equation}
Hence, with $S(t)=\sum_i\|\bm v_i(t)\|$,
\begin{equation}
 |H_{\rm ref}(T)-H_{\rm ref}(0)|\le\int_0^T e(t)S(t)\,dt.
 \label{eq:work-bound}
\end{equation}
Model changes at fixed positions and velocities leave $H_{\rm ref}$ continuous, so the identity extends piecewise between updates. Thermostats, velocity rescaling or discrete integration add their own terms. The exact oscillator experiment isolates residual-force work from these contributions.

If a force-controlled interval also has $S(t)\le S_{\max}$ and the envelope $e(t_0+\tau)\le e_0+L\tau$, a sufficient bound on its absolute residual work is
\begin{equation}
 |W(\tau)|\le S_{\max}(e_0\tau+L\tau^2/2).
 \label{eq:work-horizon}
\end{equation}
A prescribed interval work budget can therefore shorten the force-based horizon. The force envelope and speed bound together limit the absolute residual work over the interval.

\subsection{Curvature work after a reference-force correction}
Earlier on-the-fly learning work distinguishes force accuracy from energy conservation in evolving force fields~\cite{li2015fly}. In the controlled model used here, the work after a reference correction can be calculated exactly. Let $U_b$ be a fixed base potential, $\Delta U=U_{\rm ref}-U_b$, and $a$ the current reference anchor. Apply a constant force correction,
\begin{equation}
 \bm F_a(X)= -\nabla U_b(X)-\nabla\Delta U(a)+\bm r_a,
 \label{eq:anchor-correction}
\end{equation}
where $\bm r_a$ is the residual left at $a$. Hold this correction fixed until the next anchor $b$. Integration of Eq.~\eqref{eq:work-identity} then gives exactly
\begin{align}
 W_{a\to b}&=\bm r_a\cdot(b-a)+D_{\Delta U}(b,a),\label{eq:curvature-work}\\
 D_{\Delta U}(b,a)&=\Delta U(b)-\Delta U(a)
                   -\nabla\Delta U(a)\cdot(b-a).\nonumber
\end{align}
For convex $\Delta U$, the remainder equals the Bregman divergence~\cite{bregman1967}. For $\Delta U\in C^2$ on a region containing the straight segment from $a$ to $b$, writing $d=b-a$ yields
\begin{equation}
 D_{\Delta U}(b,a)=\int_0^1(1-s)d^{\mathsf T}
 H_\Delta(a+sd)d\,ds.
 \label{eq:curvature-remainder}
\end{equation}
Consequently, exact matching at the anchor ($\bm r_a=0$) leaves nonnegative segment work when the reference-minus-base curvature is positive semidefinite, and nonpositive work when it is negative semidefinite. An indefinite mismatch has no universal sign. If its spectral norm is bounded by $K_\Delta$ along that segment,
\begin{equation}
 |W_{a\to b}|\le\|\bm r_a\|_2\|d\|_2
                     +\tfrac12K_\Delta\|d\|_2^2.
 \label{eq:curvature-work-bound}
\end{equation}
For this fixed base potential and constant correction, prospective displacement and curvature bounds would convert Eq.~\eqref{eq:curvature-work-bound} into a work-based horizon.

There is also a conditional lower bound on the number of corrected segments. If every anchor is matched exactly and $H_\Delta\succeq\mu I$ with $\mu>0$ on all anchor-to-anchor straight segments, then for $M$ propagated segments and $\ell_a=\sum_{j=1}^{M}\|d_j\|_2$,
\begin{equation}
 W_{\rm tot}\ge\frac\mu2\sum_{j=1}^{M}\|d_j\|_2^2
            \ge\frac{\mu\ell_a^2}{2M}.
 \label{eq:work-reference-floor}
\end{equation}
The second inequality is Cauchy--Schwarz. For $W_{\max}>0$, keeping $W_{\rm tot}\le W_{\max}$ requires $M\ge\mu\ell_a^2/(2W_{\max})$ under these assumptions. Here $M$ counts exactly corrected propagation segments and $\ell_a$ is the polygonal length through their anchors. Both depend on the realized partition and dynamics; a closed excursion can have zero anchor displacement.

In the oscillator, $\Delta U=(k-1)x^2/2$, so the segment work is $r_a(b-a)+(k-1)(b-a)^2/2$. Across all stored segments in the 16 within-class horizon trajectories, the mean final reference-energy change is $+51.934\%$. Anchor residual work contributes $+15.421\%$ and the curvature term contributes $+36.513\%$. Each contribution is normalized by that trajectory's initial reference energy before averaging. The exact segment identity closes to $1.5\times10^{-16}$ reduced energy units; the full work--energy balance closes to $2.2\times10^{-13}$ across all 80 archived trajectories. Independent closed-form checks with positive, zero and negative mismatch verify the predicted signs, including repeated exact force matching. This decomposition uses the existing reference calculations and stochastic trajectories. Model changes leave positions, momenta and $H_{\rm ref}$ continuous; work accumulates during subsequent motion, not as an instantaneous energy jump at retraining.

\subsection{Local work at the force-budget boundary}
\label{app:local-budget-work}
After exact force matching, the force budget is locally related to the work that accumulates before the budget is reached. Set the local clock to zero at an exactly matched anchor $a$, let $\bm v_0=\dot X(0)\ne0$, and retain the same base potential and correction. Assume that $H_\Delta$ is locally Lipschitz and that $\dot X$ is locally Lipschitz. Taylor expansion of Eq.~\eqref{eq:curvature-work} gives
\begin{align}
 \bm r(t)&=H_\Delta(a)\bm v_0t+O(t^2),\nonumber\\
 e(t)&=g_0t+O(t^2),\qquad
 g_0=\|H_\Delta(a)\bm v_0\|_{\infty,2}>0,\nonumber\\
 W(t)&=\tfrac12\bm v_0^{\mathsf T}H_\Delta(a)\bm v_0t^2+O(t^3),
 \label{eq:local-time-work}
\end{align}
where $\|\bm z\|_{\infty,2}=\max_i\|\bm z_i\|_2$. Define the first time at the budget by $t_{\eps}=\inf\{t>0:e(t)\ge\eps\}$. For sufficiently small $\eps$, continuity and $|e(t)-g_0t|\le Bt^2$ imply $t_{\eps}\le2\eps/g_0$ and $|t_{\eps}-\eps/g_0|\le4B\eps^2/g_0^3$. Thus
\begin{align}
 t_{\eps}&=\frac{\eps}{g_0}+O(\eps^2),\nonumber\\
 W(t_{\eps})&=\frac{\eps^2}{2g_0^2}
       \bm v_0^{\mathsf T}H_\Delta(a)\bm v_0+O(\eps^3).
 \label{eq:local-budget-work}
\end{align}
The norm inequalities and continuity also cover ties and changes in the atom attaining the maximum. Local derivative bounds, the trajectory and nonzero $g_0$ determine the remainder constants.

The directional residual-work coefficient $\mathcal C_{\rm rw}$ in Eq.~\eqref{eq:main-work-coefficient} separates spatial distribution from signed curvature. For the unit direction $\bm u=\bm v_0/\|\bm v_0\|_2$ in full configuration space and the directional residual response $\bm h_u=H_\Delta(a)\bm u$, define
\begin{align}
 n_{\rm eff}&=\frac{\sum_i\|\bm h_{u,i}\|_2^2}
                         {\max_i\|\bm h_{u,i}\|_2^2}\in[1,n_{\rm at}],\nonumber\\
 \rho_u&=\frac{\bm u\cdot\bm h_u}{\|\bm h_u\|_2^2},\qquad
 W(t_{\eps})=\tfrac12\eps^2n_{\rm eff}\rho_u+O(\eps^3).
 \label{eq:response-participation-work}
\end{align}
The factorization gives $\mathcal C_{\rm rw}(\bm u)=n_{\rm eff}\rho_u$. Here $n_{\rm eff}$ measures the squared residual response relative to its largest atomic contribution; it is not the usual inverse participation ratio. The signed inverse-curvature quantity $\rho_u$ has units of length squared per energy. Both factors concern the error potential, rather than the material's mass-weighted dynamical matrix. In the illustrative case $H_\Delta=kI$, $k\ne0$, the leading work is $\eps^2/[2k\max_i\|\bm u_i\|_2^2]$. At the same maximum-atom force budget, a direction spread equally over all atoms can therefore accumulate $n_{\rm at}$ times the leading work of a direction confined to one atom in this example. In general, spatial distribution and curvature must be considered together.

The directional residual-work coefficient is unchanged under $\bm v_0\mapsto\alpha\bm v_0$ for each fixed $\alpha>0$, whereas the leading time to the budget scales as $1/\alpha$. This is a local small-budget statement; acceleration prevents interpreting it as an exact time rescaling at finite budget, and the limit is not uniform as $\alpha\to0$. An indefinite mismatch can have $\bm h_u\ne0$ but $\bm u\cdot\bm h_u=0$, giving zero leading work despite growing force error. When $\bm h_u=0$, the divisions above do not apply. Fixed reference noise and finite sampling intervals also preclude identifying the small-budget limit from arbitrarily small measured errors. Directional probes can estimate $\mathcal C_{\rm rw}$ for a subsequent test, while the sign, amplitude and resolved interval require independent reference evidence.

\subsection{Local persistence of a force--work reversal}
\label{app:directional-robustness}
At a fixed time, a strict reversal between force-error and residual-work orderings persists under sufficiently small changes in initial direction, independently of the small-budget expansion. Fix a configuration $a$, positive atomic masses $m_i$, and initial total kinetic energy $K_0>0$. Let $U_b,U_{\rm ref}\in C^2(\Omega)$ on an open configuration domain and retain the anchored potential
\begin{align*}
 U_a(X)&=U_b(X)-\bm c\cdot(X-a),\\
 \bm c&=-\nabla(U_{\rm ref}-U_b)(a).
\end{align*}
The allowed Euclidean unit velocity directions and initial velocities are
\begin{align*}
 \mathcal D&=\left\{\bm u:\|\bm u\|_2=1,\ \sum_i m_i\bm u_i=0\right\},\\
 \bm v_0(\bm u)&=\left(\frac{2K_0}{\sum_i m_i\|\bm u_i\|_2^2}\right)^{1/2}\bm u.
\end{align*}
This normalization fixes the initial kinetic energy and removes center-of-mass momentum for every direction, including systems with unequal masses.

Let $X(t;\bm u)$ solve Newton's equations under $U_a$ from $(a,\bm v_0(\bm u))$. Assume a common existence interval $[0,T]$ in $\Omega$ for directions near two choices $\bm u_1,\bm u_2\in\mathcal D$. With $\Phi=U_{\rm ref}-U_a$ and $\bm R=\nabla\Phi$, define
\begin{align*}
 e_T(\bm u)&=\max_i\|\bm R_i(X(T;\bm u))\|_2,\\
 W_T(\bm u)&=\int_0^T\bm R(X(t;\bm u))\cdot\dot X(t;\bm u)\,dt\\
 &=\Phi(X(T;\bm u))-\Phi(a).
\end{align*}
If the two directions have strict gaps
\begin{align*}
 \delta_e&=e_T(\bm u_2)-e_T(\bm u_1)>0,\\
 \delta_W&=W_T(\bm u_1)-W_T(\bm u_2)>0,
\end{align*}
then both inequalities hold for every pair in sufficiently small relative open neighborhoods of $\bm u_1$ and $\bm u_2$ in $\mathcal D$.

\textit{Proof.} The velocity normalization is smooth, and the Newtonian flow depends continuously on its initial data. Each atomic residual norm is continuous, as is their finite maximum, including at ties or changes of the maximizing atom. The endpoint expression also makes $W_T$ continuous. Choose the neighborhoods so that each $e_T$ changes by less than $\delta_e/4$ and each $W_T$ by less than $\delta_W/4$. The two perturbed gaps then exceed $\delta_e/2$ and $\delta_W/2$, respectively.

Thus a strict reversal persists in an open neighborhood of the initial directions at fixed structure, initial kinetic energy and common time $T$. Equation~\eqref{eq:local-budget-work} gives the complementary comparison at direction-dependent first force-budget crossings.

\section{Controlled calculations and supplementary diagnostics}
\label{sec:controlled}
\begin{figure*}[!tp]
 \centering
 \includegraphics[width=\textwidth]{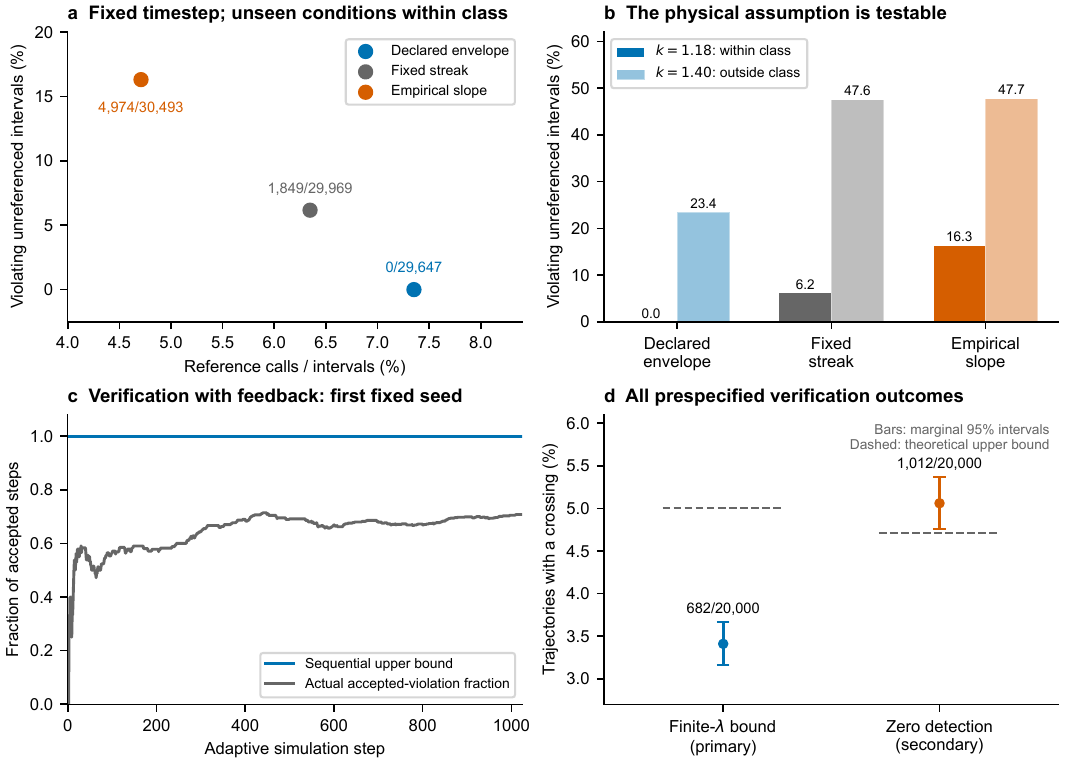}
 \caption{\label{fig:controlled}Controlled tests of prospective admission and retrospective verification. (a) Sixteen held-out oscillator trajectory pairs at fixed $\Delta t=0.01$, with the supplied mismatch bound valid. Counts above points give violating/unreferenced intervals. The physical-envelope rule uses 15.85\% more references than the fixed-streak baseline. (b) The same rule fails outside its declared stiffness class. Using the empirical derivative from calibration causes violations even within the class. (c) Actual accepted-violation fraction and sequential upper bound on a high-risk path from the first prespecified statistical simulation seed. (d) All prespecified crossing outcomes over 20,000 independent simulations. Dashed lines bound trajectory event probabilities; points and error bars give empirical frequencies and separate marginal 95\% Clopper--Pearson intervals. The finite-$\lambda$ primary frequency lies below its comparator; the zero-detection secondary sample interval lies above its comparator (post hoc one-sided binomial $p=0.01079$).}
\end{figure*}

Figure~\ref{fig:controlled} compares prospective oscillator admission with independent verification under adaptive feedback.

\subsection{Prospective admission at fixed physical timestep}
We first test whether a physical residual envelope can allocate reference calls more effectively than a fixed maximum acceptance streak. The reference is a unit-mass oscillator with $F^{\rm ref}(x)=-kx$; the surrogate is $F^{\rm sur}(x)=-x+b$, with a label-updated bias $b$. The declared physical class is $|k-1|\le D=0.2$. Between updates the surrogate flow is exact,
\begin{equation}
 x(\tau)=b+(x_0-b)\cos\tau+v_0\sin\tau.
 \label{eq:oscillator}
\end{equation}
Its speed is bounded by $v_{\max}=\sqrt{(x_0-b)^2+v_0^2}$. Thus a reference-measured initial residual $e_0$ gives
\begin{equation}
 e(\tau)\le e_0+Dv_{\max}\tau,\qquad
 \tau_*=(\eps-e_0)/(Dv_{\max})
 \label{eq:oscillator-horizon}
\end{equation}
when $e_0\le\eps$ and the denominator is positive. The supplied curvature-mismatch bound makes this an analytical test of prospective force control.

Every policy advances 2000 intervals of length $\Delta t=0.01$ in reduced physical units, with force tolerance 0.05. Exact propagation of each surrogate interval removes timestep integration error. A reference call updates the bias by 80\% of the observed discrepancy; if the remaining residual is at least the tolerance, the same label supplies a full bias correction. All subsequent trajectories are generated forward from their own states. An additional reference check with probability 0.02 may interrupt an eligible streak and corrects the force before the next interval. The primary outcome of this horizon experiment is therefore the maximum error throughout each \emph{unreferenced interval}. The separate verification experiment below measures the pre-check accepted-force count in Proposition~2.

The entire protocol was fixed locally before execution. Eight trajectories at $k=1.10$ determine the reference cost used to select a fixed-streak baseline, $K=18$, without consulting its violations. Two calibration force calls also give the empirical derivative estimate $\widehat D=0.10$. Both quantities are frozen before testing 16 new initial-condition seeds and amplitudes at $k=1.18$, within the declared class but outside the derivative-calibration condition. The baseline comparison permits a prespecified 20\% relative difference in reference cost. A further eight seeds at $k=1.40$ test behavior outside the declared class. No timestep or parameter is adjusted after inspecting these results.

On the 16 held-out trajectories, the declared-envelope rule has zero violating intervals among 29,647 unreferenced intervals, using 2353 reference calls out of 32,000 intervals. The fixed-streak rule has 1849 violations among 29,969 unreferenced intervals (6.17\%), using 2031 reference calls. Reference cost is 15.85\% higher for the envelope rule, within the specified tolerance. The envelope leaves 92.65\% of intervals unreferenced. Counts summarize 16 independent trajectory pairs, with correlated intervals within each trajectory.

The two boundary tests change the residual-growth estimate or the oscillator stiffness. Replacing the declared bound by the frozen empirical derivative gives 4974 violations among 30,493 unreferenced intervals (16.31\%) at the held-out stiffness. At $k=1.40$, outside the declared class, the nominal envelope rule has 3384 violations among 14,451 unreferenced intervals (23.42\%). Force control therefore depends on a valid bound for the declared stiffness class.

The shared development cost is 1015 reference evaluations; including it gives 3368 evaluations for the held-out envelope policy and 3046 for the fixed-streak baseline. Measuring all errors and evaluating the exact-reference trajectories require additional calculations, whose results are excluded from the admission decisions. Costs in this analytical-force experiment are measured as reference-call counts.

The within-class envelope trajectories accumulate a mean final change of $+51.93\%$ in the \emph{reference} Hamiltonian over the 20-unit duration. Exact interval propagation isolates the residual work responsible for this change. Appendix~\ref{sec:work} separates the contribution of curvature mismatch from that of the residual remaining at each anchor.

\subsection{Independent verification with feedback}
A binary stress test implements Proposition~2 under adaptive risk and admission probabilities. A detected violation reduces future risk and admission for a cooldown period; feedback uses only observed violations. Each acceptance, including a checked one, retains its original surrogate outcome.

A single study fixed 20,000 independent seeds, 1024 steps per seed, $p=0.10$, $\lambda=\log2$, and $\eta=0.05$ before sampling. The primary outcome is whether the bound fails \emph{at any time} within a trajectory. Failures occur in 682 of 20,000 trajectories, or 3.410\%, with a marginal 95\% Clopper--Pearson interval of 3.163--3.671\%, below the theoretical upper bound of 5\%. These intervals use the independent simulation replicates, not the dependent time steps. Exact rational checks of the conditional multiplier and a fully enumerated depth-10 adaptive reference-check tree also preserve the martingale mean of one.

The mean terminal accepted-violation fraction in this high-risk process is 74.499\%, and the mean capped upper bound is 99.212\% (Fig.~\ref{fig:controlled}c). The test measures simultaneous coverage; the gap shows how far the upper bound lies above the actual fraction in this process.

The prespecified secondary outcome probes the sharp zero-detection limit. For $p=0.10$, the event $E_0=\{\exists t\le1024:D_t=0,V_t\ge29\}$ has probability at most $B=0.9^{29}=4.710\%$. It occurs in 1012 of 20,000 simulations (5.060\%; marginal 95\% interval 4.760--5.373\%), placing this sample interval above the comparator. Independent reconstruction using the original pseudorandom streams reproduces all 20,000 event-prefix counts and first-event times. The sample excess has a post hoc binomial tail probability of 0.01079. The original seeds, parameters and outcomes are retained.

An independent deterministic calculation evaluates the protocol's finite-time probability. Absorbing probability propagation on the integer state lattice gives $\Prob(E_0)=0.04710128697246244$ in float64. An analytical check divides the undetected path mass by $0.9^V$; the resulting normalized process reaches saturated admission and risk probabilities by step 113. Its remaining 911 steps each have conditional violation probability at least $98/125$. Hence
\begin{equation}
 0\le B-\Prob(E_0)\le
 B\sum_{j=0}^{28}\binom{911}{j}
 (98/125)^j(27/125)^{911-j}.
 \label{eq:finite-thinning}
\end{equation}
The right-hand side bounds the gap $B-\Prob(E_0)$ below $1.741\times10^{-539}$ for the ideal integer-state protocol. This analytical result shows that the event probability essentially attains $B$. The finite-precision calculation has a conservative recursion-roundoff bound of $1.46\times10^{-11}$, and the probability difference from the original floating-state protocol is bounded by $3.73\times10^{-9}$. The original Monte Carlo estimate is 0.3499 percentage points above this nearly attained comparator; the deterministic probability and the finite-sample excess are distinct results.

\section{Prospective molecular protocol}
\label{app:forward-molecular}
The molecular trajectories use the same four-member MACE architecture and PBE/def2-SVP reference as the archived water study. A common checkpoint is fitted on archived configurations 0--63, and configurations 64--95 supply initial residual-calibration scores. Frozen-model replay on configurations 96--301 selects reference period 16 from $1,2,4,8,16$ and calibrated-bound multiplier 1 from $1,1.5,2$, before new forward labels are obtained. Initial geometries are archived configurations 0 and 151. Gaussian momenta are mass weighted, stripped of center-of-mass motion, and scaled to $6k_B(300\,\mathrm K)/2$ kinetic energy before the paired velocity factors are applied. Velocity seeds 2026090501 and 2026090502 define new directions at these known geometries.

For current-model residual vectors $\bm r_j$ at the four most recent reference geometries, the empirical coefficient is
\begin{equation}
 \widehat\kappa=2\max_j
 \frac{\|\bm r_j-\bm r_{j-1}\|_{\infty,2}}
      {\|\widetilde X_j-\widetilde X_{j-1}\|_2},
 \label{eq:empirical-path-slope}
\end{equation}
where $\widetilde X$ removes center-of-mass translation and pairs closer than $10^{-4}$~\AA\ are excluded. If no pair is usable, the rule supplies no positive horizon. All residuals use the same current model. Starting from a reference-measured $e_0$, frozen-model velocity-Verlet forecasts supply a cumulative configuration-path length $\ell(\tau)$. The empirical rule chooses the last 0.5-fs grid point satisfying $e_0+\widehat\kappa\ell(\tau)\le0.10$~eV/\AA, up to an 8-fs forecast cap. A zero result refuses the first grid point.

The six trajectories with varied initial velocities keep the model weights fixed and advance to 9.5 fs. Initial labels are reused, and all 114 future force-evaluation states receive hidden reference labels after the horizon prediction is written. Eight subsequent central-difference probes use displacements $\pm h\dot X/\|\dot X\|_2$, with $h=0.001$ and $0.002$~\AA, to measure the local derivatives. The source, common checkpoint and saved predictions record the fixed protocol. Full-step momenta satisfy both velocity-Verlet identities to $1.2\times10^{-16}$ ASE momentum units and the saved position precision. Force comparisons use the discrete evaluation grid.

\subsection{High-velocity forward boundary test}
\label{app:hot-forward}
Development reuses the two labelled $2\times$ velocity paths. We replayed periods $K\in\{1,2,4,8,16\}$ and calibrated multipliers $\{1,1.5,2,3,4,8,16,10^6\}$, resetting the 32 calibration scores for each candidate and revealing only refused or independently checked labels. Candidates accepting at least four of 20 states per path all exceeded the 5\% accepted-violation criterion on at least one path. The prescribed fallback minimized the worst-path violation fraction, then reference demand including checks, selecting $K=8$ and multiplier 1. The all-reference policy covers $K=1$.

The settings were fixed before generating paths from velocity seeds 2026090521 and 2026090522 at the same two known geometries, with initial momenta doubled relative to the 300-K kinetic normalization. Each of four policies generated 40 states spanning 19.5 fs at 0.5-fs spacing. Model weights remained fixed; the horizon used the selected spread multiplier and Eq.~\eqref{eq:empirical-path-slope}. On refusal the reference force propagated the step. Each accepted proposal preceded an independent $p=0.1$ check and retained its proposed force after checking. Labels updated future scores and anchors; hidden measurement labels were excluded from control. All eight paths and 312 future reference evaluations completed (Table~\ref{tab:hot-forward}).

\begin{table}[!htbp]
\caption{\label{tab:hot-forward}Complete high-velocity boundary test. Each pair of entries refers to the first and second new velocity seed, respectively. $N$ counts accepted surrogate forces, $V$ their violations, and $N_{\rm ref}$ all policy requests including independent checks. Energy drift is in meV per molecule. All-reference and zero-acceptance policies have undefined accepted-violation fractions.}
\begin{ruledtabular}
\begin{tabular}{lrrrr}
Policy & $N$ & $V$ & $N_{\rm ref}$ & $\max|\Delta H_{\rm ref}|$\\
Reference & 0, 0 & 0, 0 & 40, 40 & 2.498, 3.210\\
Periodic & 35, 35 & 21, 19 & 11, 9 & 13.509, 11.782\\
Calibrated & 3, 1 & 0, 1 & 37, 40 & 1.698, 6.113\\
Horizon & 0, 0 & 0, 0 & 40, 40 & 2.498, 3.210
\end{tabular}
\end{ruledtabular}
\end{table}

The horizon refused every surrogate force and reproduced the reference paths; calibration requested references at nearly every state. On the reference paths, model errors exceeded the budget at 23/40 and 21/40 states. Finite-timestep reference-energy drift is included in Table~\ref{tab:hot-forward}. The adapted-model comparison below tests propagation after reusing these labels.

\subsection{Label reuse and replicated molecular propagation}
\label{app:molecular-feasibility}
\begin{figure*}[!tp]
\centering
\includegraphics[width=\textwidth]{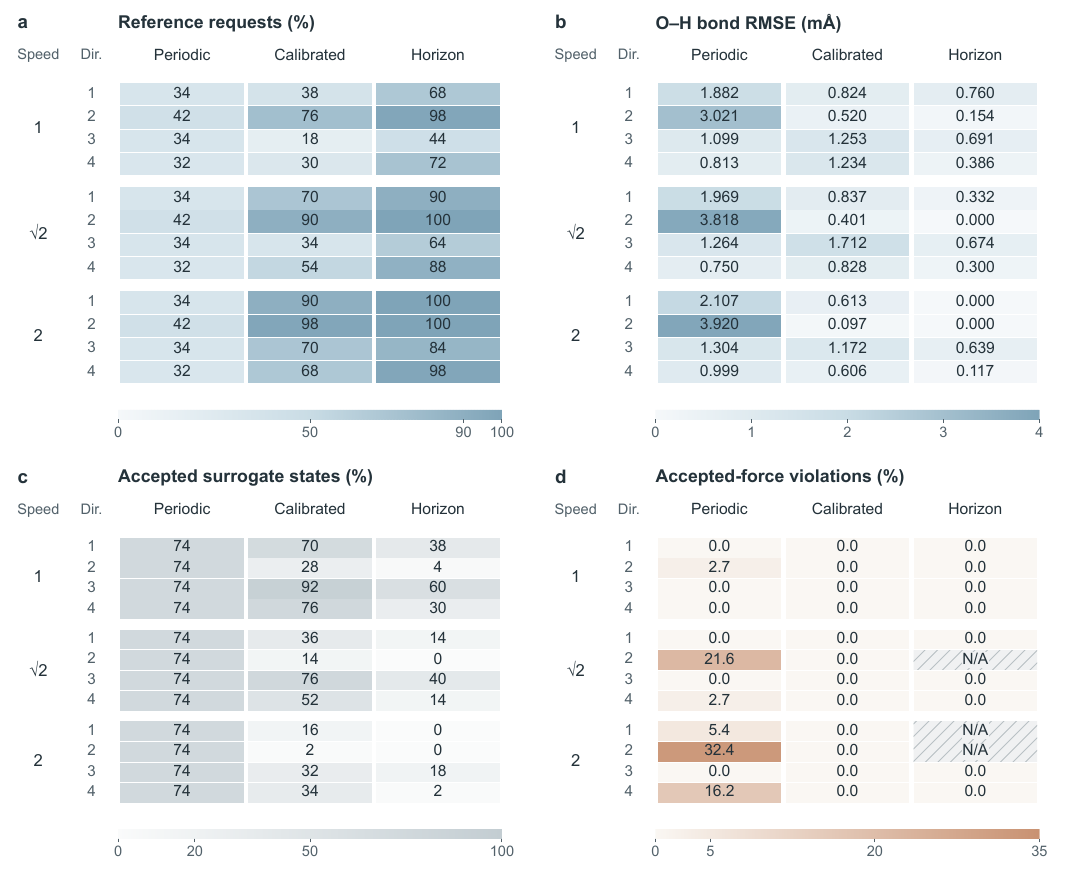}
\caption{\label{fig:molecular-feasibility}\textbf{Reference demand and structural fidelity in paired molecular propagation.} All 36 approximate-force trajectories are shown. Columns identify policies; row groups identify initial speed factors $1$, $\sqrt{2}$ and $2$, each with four fresh velocity directions at two known geometries. Each cell reports one complete 50-state trajectory (24.5 fs); each panel uses a fixed linear color scale across all policies and speeds. \textbf{a,} Reference requests as a percentage of 50 states, including independent checks. \textbf{b,} O--H bond root-mean-square error (RMSE) over both bonds and all 50 states against the reference trajectory with the same initial direction and speed. \textbf{c,} Percentage of states accepted for surrogate forces. \textbf{d,} Observed force-budget violations among accepted states, in percent. Hatched N/A cells mark the three horizon paths with no acceptances, for which the accepted-violation fraction is undefined. Their zero RMSE in b results from reference forces being used throughout. Screening required reference requests $\leq90\%$, acceptance $\geq20\%$ and observed accepted violations $\leq5\%$ in every direction at a speed. At the original speed, the horizon reduces O--H RMSE relative to calibration in all four pairs while requesting more references. The 12 reference paths are included among all 48 paths in the source data. They have 100\% reference requests, zero acceptances, undefined accepted-violation fractions and zero self-deviation.}
\end{figure*}

We adapted the foundation readouts once using the high-velocity labels, the original four-member recipe, 50 epochs and a fixed seed. Training combines archived configurations 0--23 with the first new reference path, giving 64 entries and 62 unique geometries. The second path supplies 40 validation geometries, and the original 32 cold calibration geometries are also disjoint from training. The fit is then frozen for the forward comparison.

With the adapted model held fixed, four new random seeds generated velocity directions at known archived geometries 0, 151, 0 and 151; the seeds are retained with the source data. Each direction was paired across speed factors $1$, $\sqrt{2}$ and $2$, and across reference, periodic, calibrated and horizon policies. Each policy generated its own 50-state trajectory, spanning 24.5 fs at 0.5-fs spacing. The initial kinetic energies correspond to 300, 600 and 1200 K. The independent units are the four fresh velocity directions, paired across speeds and policies.

The force budget, check probability, empirical growth prefactor and forecast cap remained 0.10 eV/\AA, 0.1, 2 and 8 fs. The initial 32 calibration scores were recomputed with the adapted model. Replay of the two earlier high-velocity development paths selected calibrated multiplier 1. The best nontrivial periodic candidate was $K=4$, which again failed the development risk criterion; $K=1$ is covered by the reference policy. Before generating new trajectories, we fixed a practical screening criterion for every direction at a tested speed. Each trajectory had to accept at least 20\% of states, have at most 5\% observed accepted violations and request at most 90\% as many references as full reference propagation. These thresholds assess reference demand together with acceptance and observed violations. Frozen proposals, independent checks, force retention and hidden measurement labels follow the preceding protocol.

All 48 trajectories completed with 2352 new reference evaluations, all of which converged; the 48 initial labels were reused. The independent reconstruction of forces, decisions, checks and velocity-Verlet updates passed for every path. Figure~\ref{fig:molecular-feasibility} shows all approximate-force paths. At the original speed, the horizon accepted $(19,2,30,15)$ states and requested $(34,49,22,36)$ references; calibration accepted $(35,14,46,38)$ and requested $(19,38,9,15)$. Neither policy had an observed accepted violation. Compared with calibration, the horizon lowered O--H bond and angle RMSE against the paired reference trajectories in every pair, but used more references in every pair. Its second direction failed the practical screening criterion. At factors $\sqrt{2}$ and $2$, horizon acceptances fell to $(7,0,20,7)$ and $(0,0,9,1)$, with reference requests $(45,50,32,44)$ and $(50,50,42,49)$. No speed level passed the four-direction horizon criterion. Calibration passed at factor 1, and also had no observed accepted violations at the other two factors. The periodic policy had $(0,1,0,0)$, $(0,8,0,1)$ and $(2,12,0,6)$ violations among 37 acceptances per path at the three factors, respectively.

These trajectories measure structural fidelity and reference demand together. Maximum energy drift in reference propagation spans 0.466--6.271 meV from finite-timestep effects. Complete hidden labeling determines the reported accepted-violation counts; the sequential upper bound in Eq.~\eqref{eq:main-verification} is 100\% for each nonempty individual acceptance record.

\section{Electronic-reference sensitivity}
\label{sec:scope}
The exchange--correlation approximation, basis, pseudopotential, electronic smearing and convergence tolerances specify the reference force. We measure how this force changes with three numerical settings at a fixed interface configuration.

Increasing the plane-wave cutoff from 60 to 70 Ry changes forces by RMS 0.0032 eV/\AA\ and maximum 0.016 eV/\AA. Changing Fermi--Dirac smearing from 0.02 to 0.01 Ry changes them by RMS 0.058 and maximum 0.389 eV/\AA. Changing $\Gamma$ sampling to $2\times2\times1$ gives RMS 0.015 and maximum 0.083 eV/\AA. These paired calculations measure the sensitivity of the specified reference force; comparison with the force budget uses the maximum-atom values.

\section{Decision and verification protocol}
\label{app:protocol}
At each force evaluation: (i) freeze the current configuration and surrogate force; (ii) compute admission from past labels and current inexpensive quantities; (iii) on refusal use the reference force, and on acceptance draw an independent Bernoulli check; (iv) if checked, compare the reference with the force frozen in step (i); (v) retain every pre-check acceptance in $N_t$ and every detected violation in $D_t$; (vi) propagate an accepted step with its frozen force, then update later decisions and models. A failed reference calculation leaves its check unresolved and requires completion or conservative handling.

\section{Empirical residual calibration}
\label{app:calibration}
For a fixed predictor with exchangeable scores, the split-conformal upper quantile uses rank $j=\lceil(n+1)(1-\alpha)\rceil$ in a sample augmented by $+\infty$. At $\alpha=0.05$, a finite distribution-free quantile requires at least 19 scores. The historical implementation instead used a minimum window of 16 and rank clipped to $n$, with scores collected under evolving labels and model parameters. Its admission score is therefore empirical. Proposition~2 supplies an independent accepted-error measurement protocol.

The water committee uses four readouts, a window of 64 normalized residuals, $\delta=10^{-3}$ eV/\AA, and nominal $\alpha=0.05$. Fine-tuning is triggered every eight revealed labels, using Adam with learning rate $10^{-3}$ for 50 epochs and loss $\mathrm{MSE}(E)/n_{\rm at}+10\mathrm{MSE}(\bm F)$. Every water frame is evaluated at the same PySCF RKS PBE/def2-SVP reference level with SCF tolerance $10^{-9}$. Complete labeling determines the finite-stream violation count on these correlated configurations.

The budget sets reference demand on this stream. At $\eps=0.005$ eV/\AA, all 302 configurations are refused and the accepted-violation fraction is undefined. At 0.02 eV/\AA, 291 references leave 11 acceptances, all within budget. At 0.10 eV/\AA, 78 references leave 224 acceptances, also within budget. Each operating point has its own policy-specific training history.

\section{Interface reference and trajectory settings}
The cell measures approximately $10.53\times10.53\times52.15$ \AA$^3$, with electrolyte preparation density 1.28 g/cm$^3$. The source trajectory uses velocity Verlet at 0.5 fs and 300-K initial velocities. It supplies the two prospective origins and the archived coordinates and forces illustrated in Fig.~\ref{fig:materials-principle}. The committee combines MACE-MP-0b3-medium and MACE-MPA-0-medium. Reference calculations use PWSCF 7.5~\cite{giannozzi2017advanced}, PBE plus D3~\cite{grimme2010d3}, pslibrary kjpaw PAW datasets, $\Gamma$ sampling, 60/600-Ry wavefunction/charge-density cutoffs, Fermi--Dirac smearing 0.02 Ry, and 942 bands for 1584 electrons. Historical SCF settings use local-TF mixing with coefficient 0.20 and history 12, Davidson subspace 8, convergence threshold $10^{-6}$ Ry and density reuse. The prospective calculations use the tighter settings below. Appendix~\ref{sec:scope} reports reference-force sensitivities.

\subsection{Prospective directional tests at the interface}
\label{app:interface-prospective}
We test whether local residual derivatives predict force error and residual work during subsequent motion of the same 474-atom interface. Two saved configurations, steps 36 and 161, provide the reference origins. Each origin has two fresh Gaussian momentum directions, using seeds 2026090561--2026090564 in that order. Center-of-mass momentum is removed and the initial kinetic energy is scaled to $(3n_{\rm at}-3)k_B(600\,\mathrm K)/2$. The four trajectories thus vary the initial momenta at two fixed structures. The base committee is frozen, and each trajectory retains the constant correction $\bm c=\bm F_{\rm ref}(o)-\bm F_b(o)$ throughout a 4-fs velocity-Verlet segment. The primary and comparison timesteps are 0.125 and 0.0625 fs, with identical initial momenta, no thermostat and no model updates. Positions remain on a common unwrapped local branch. The resulting anchored potential is $U_a(X)=U_b(X)-\bm c\cdot(X-o)$.

Development uses the two reference origins and four displacements per direction, $\pm0.02\bm u$ and $\pm0.04\bm u$~\AA, where $\bm u$ is the Euclidean unit vector along the full initial configuration velocity. For residual $\bm R=\bm F_b+\bm c-\bm F_{\rm ref}$, define $\bm q_h=[\bm R(o+h\bm u)-\bm R(o-h\bm u)]/(2h)$ and retain $\widehat{\bm q}=\bm q_{0.04}$. Displacements at the saved precision are used in energy differences and finite-displacement defects. The directional curvature estimate $C=\bm u\cdot\widehat{\bm q}$ gives the leading work prediction, while $\|\widehat{\bm q}\|_{\infty,2}$ gives the leading maximum-atom residual growth. The ratio $\widehat{\mathcal C}_{\rm rw}=C/\|\widehat{\bm q}\|_{\infty,2}^2$ estimates the directional residual-work coefficient in Eq.~\eqref{eq:main-work-coefficient} from these finite probes.

For actual displacement $\bm d=X-o$, write $\alpha=\bm u\cdot\bm d$ and $\bm z=\bm d-\alpha\bm u$. The frozen empirical prediction is
\begin{align}
 \bm P(X)&=\alpha\widehat{\bm q},\nonumber\\
 B(X)&=\|\bm P(X)\|_{\infty,2}+\rho(X),\label{eq:interface-empirical-envelope}\\
 \rho(X)&=\rho_0+\eta_\parallel|\alpha|+K_\perp\|\bm z\|_2
                    +\tfrac12M\|\bm d\|_2^2.\nonumber
\end{align}
Here $\rho_0=0.00025$ eV/\AA, $\eta_\parallel=2\|\bm q_{0.02}-\bm q_{0.04}\|_{\infty,2}$, and $K_\perp$ is twice the largest derivative norm over both directions and scales at the same origin. For each direction, $M$ is four times the largest development value of $\|\bm R-\alpha\widehat{\bm q}\|_{\infty,2}/\|\bm d\|_2^2$. These coefficients estimate the physical envelope from a finite set of probes. We test their predictions on future configurations beyond the development displacements.

The primary force budget is 0.25 eV/\AA. The forecast domain is $t\le1$ fs and $\|\bm z\|_2\le0.1\|\bm d\|_2$. Admission requires every primary integration state up to that time to lie in this domain and satisfy $B\le\eps$; the first failure ends the accepted prefix. The four predicted windows are 0.375, 0.25, 0.25 and 0.375 fs. Auxiliary budgets 0.05, 0.10 and 1.0 eV/\AA\ use the same score. The reference-evaluation grid is 0.125, 0.25, 0.375, 0.5, 0.75, 1, 1.5, 2, 3 and 4 fs for each direction. All predictions, windows and reference geometries were fixed before these 40 future evaluations were submitted.

We assess the accuracy of the directional residual prediction separately from envelope coverage. From candidate times 0.125, 0.25, 0.375 and 0.5 fs, the diagnostic set retains those with predicted amplitude $\|\bm P\|_{\infty,2}\ge3\nu$. The development sensitivity scale $\nu$ is 0.007149 eV/\AA\ at the first origin and 0.006579 eV/\AA\ at the second, obtained from the change between the archived and tightened origin forces. On the fixed diagnostic set, the declared test requires each measured amplitude to be at least $3\nu$, each measured-to-predicted amplitude ratio to lie in $[0.5,2]$, and
\begin{equation}
 S_{\rm dir}=\frac{\sum_t\|\bm R_t-\bm P_t\|_{\infty,2}^2}
                       {\sum_t\|\bm R_t\|_{\infty,2}^2}\le0.5.
\end{equation}
The first direction at each origin is the primary growth test; all four directions are reported. The vector-radius comparison $\|\bm R-\bm P\|_{\infty,2}/\rho$ and scalar ratio $e/B$ provide additional diagnostics.

Reference settings retain the interface PBE-D3 approximation, cutoffs, sampling and fixed Fermi--Dirac broadening above. The developmental and future labels use an SCF threshold of $10^{-8}$ Ry, atomic-potential initialization and full-accuracy diagonalization. The potential $U_{\rm ref}$ is the printed variational free energy $E-TS$, whose derivatives give the forces at the fixed electronic broadening. The endpoint work is evaluated as $W=\Delta U(X)-\Delta U(o)+\bm c\cdot(X-o)$, with $\Delta U=U_{\rm ref}-U_b$. The reference-Hamiltonian change is separated into $W$ and the measured anchored-Hamiltonian integration drift $D_a$.

\begin{figure}[!t]
\centering
\includegraphics[width=\columnwidth]{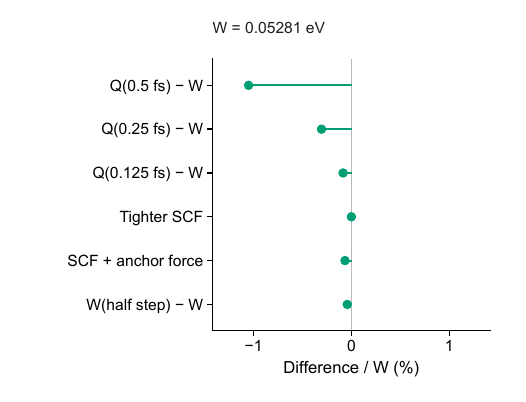}
\caption{\label{fig:prospective-numerics}\textbf{Numerical sensitivity of residual work at the p2 1-fs endpoint.}
Signed numerical differences are expressed as percentages of the endpoint work $W$; they are not statistical intervals.
The first three rows compare same-path force quadratures on 0.5-, 0.25- and 0.125-fs grids with $W$.
``Tighter SCF'' gives the fixed-correction change $\delta U(X)-\delta U(o)$; ``SCF + anchor force'' also includes $\delta\bm F(o)\cdot(X-o)$, where $\delta$ denotes the tighter reference minus the primary reference.
The final row compares half-step and primary endpoint work. The primary work criterion is satisfied, and the separate p0 half-step check is complete.}
\end{figure}

Six additional reference evaluations test whether the work at a fixed 1-fs endpoint is numerically resolved. Figure~\ref{fig:prospective-numerics} shows the resulting work differences. Selection uses only the development curvatures: for the two scales, let $\kappa_h=\|\bm v_0\|_2^2\bm u\cdot\bm q_h$ and assign score $\max\{0,\min(|\kappa_{0.02}|,|\kappa_{0.04}|)-|\kappa_{0.02}-\kappa_{0.04}|\}$ when the signs agree, and zero otherwise. The higher-scoring direction at each origin is retained, with ties resolved by its original order. This selects directions 2 and 0, counting from zero, and direction 2 has the larger score. Its origin and primary 1-fs endpoint are each recalculated at $10^{-10}$ Ry, and its primary 0.625- and 0.875-fs configurations complete a 0.125-fs force-quadrature grid. The final two labels evaluate the half-timestep 1-fs endpoints of directions 2 and 0.

The force integral uses the actual coordinate increments along the primary path,
\begin{equation}
 Q_\ell=\sum_j\tfrac12(\bm R_j+\bm R_{j+1})\cdot(X_{j+1}-X_j).
\end{equation}
The grids have spacings $\ell=0.5$, 0.25 and 0.125 fs and span the same interval. With $\delta U$ and $\delta\bm F$ denoting the changes upon SCF tightening, the fixed-correction work sensitivity is $\delta W_{\rm fixed}=\delta U(X)-\delta U(o)$. The sensitivity including the anchor-force correction is $\delta W_{\rm curv}=\delta W_{\rm fixed}+\delta\bm F(o)\cdot(X-o)$; it is reported without changing the applied correction. The prescribed numerical test requires positive endpoint, finest-quadrature, tightened and half-timestep work. The finest quadrature, fixed-correction tightening and half-timestep work must each agree with $W$ within $0.05|W|+0.0005$ eV. The signal must be at least ten times each associated absolute difference, both absolute differences between adjacent quadrature levels, $|\delta W_{\rm curv}|$, and a $10^{-5}$-eV output-precision floor. The two endpoint force sensitivities must be at most 0.0025 eV/\AA. Work is called the dominant reference-energy contribution when $|D_a|<|W|/3$. The 1-fs checks evaluate the work mechanism beyond the primary admission windows.

The prospective interface validation uses 64 single-point reference calculations: 18 development evaluations, 40 future endpoints and six numerical checks. All completed successfully, using 3258.24 allocated CPU hours in total. Each calculation was allocated 96 CPUs; the total sums the product of allocated CPUs and elapsed time for each job. The four trajectory-generation jobs and development model inference used a further 34.84 allocated CPU hours. These counts describe the new validation work; the archived references and previous model preparation remain separate.

To identify which atoms contribute to the force integral, we partition it by fixed atom identities,
\begin{equation}
 Q_G=\sum_j\sum_{i\in G}\tfrac12(\bm R_{i,j}+\bm R_{i,j+1})
                    \cdot(\bm X_{i,j+1}-\bm X_{i,j}).
\end{equation}
The zero-based groups retain their initial identities: slab Li, 0--35; EC, 36--245; DMC, 246--449; and salt, 450--473. Summing the four groups recovers the complete force integral. Individual atomic and group contributions retain their signs, and their net fractions use this same integral as denominator. This decomposition is computed after endpoint evaluation by integrating each atom's residual force along its displacement. Source data retain every atomic contribution and all four early-time comparisons between directions at a common origin.

\subsection{Complete prescribed endpoint record}
\label{app:interface-record}
Table~\ref{tab:interface-endpoints} reports the four fixed paths at every prescribed reference time. The same initial-force correction defines all residuals and endpoint work on each path.
\begin{table*}[tp]
\caption{\label{tab:interface-endpoints}\textbf{Complete prescribed endpoint record for the interface.} Paths p0/p1 start from origin 36 and p2/p3 from origin 161. Stars mark evaluation times inside the frozen 0.25-eV/\AA\ admission windows. Residual work uses the endpoint energy expression. The 1.5--4-fs points are continuation probes beyond the declared 1-fs envelope domain. }
\centering\small\setlength{\tabcolsep}{3pt}
\begin{minipage}[t]{0.49\textwidth}\centering
Maximum atomic residual $e$ (eV/\AA)\par\smallskip
\begin{tabular}{rrrrr}\toprule
$t$ (fs) & p0 & p1 & p2 & p3\\\midrule
0.125 & 0.0303$^{*}$ & 0.0334$^{*}$ & 0.0137$^{*}$ & 0.0165$^{*}$\\
0.25 & 0.0600$^{*}$ & 0.0670$^{*}$ & 0.0275$^{*}$ & 0.0328$^{*}$\\
0.375 & 0.0892$^{*}$ & 0.1005 & 0.0412 & 0.0488$^{*}$\\
0.5 & 0.1177 & 0.1340 & 0.0561 & 0.0647\\
0.75 & 0.1727 & 0.2002 & 0.0880 & 0.0958\\
1 & 0.2247 & 0.2648 & 0.1223 & 0.1260\\
1.5 & 0.3191 & 0.3857 & 0.1959 & 0.1837\\
2 & 0.4000 & 0.4922 & 0.2732 & 0.2373\\
3 & 0.5170 & 0.7368 & 0.4223 & 0.3426\\
4 & 0.5708 & 0.9449 & 0.5430 & 0.4439\\
\bottomrule\end{tabular}
\end{minipage}
\hfill
\begin{minipage}[t]{0.49\textwidth}\centering
Residual work $W$ (meV)\par\smallskip
\begin{tabular}{rrrrr}\toprule
$t$ (fs) & p0 & p1 & p2 & p3\\\midrule
0.125 & 0.646 & 0.419 & 0.877 & 0.761\\
0.25 & 2.562 & 1.636 & 3.498 & 3.054\\
0.375 & 5.740 & 3.616 & 7.834 & 6.851\\
0.5 & 10.154 & 6.343 & 13.827 & 12.103\\
0.75 & 22.664 & 13.990 & 30.491 & 26.686\\
1 & 39.890 & 24.492 & 52.809 & 46.124\\
1.5 & 87.796 & 54.187 & 111.695 & 97.059\\
2 & 152.502 & 96.436 & 185.923 & 160.980\\
3 & 324.377 & 222.792 & 364.019 & 320.740\\
4 & 539.281 & 409.293 & 553.414 & 516.425\\
\bottomrule\end{tabular}
\end{minipage}
\end{table*}

Table~\ref{tab:interface-endpoints} shows the prescribed continuation probes beyond the local prediction interval. At 4 fs, the transverse displacement fraction is 0.447--0.517, above the declared envelope-domain threshold of 0.1. Residual work exceeds the frozen directional predictions by 53.1\%, 70.9\%, 17.8\% and 22.0\% for p0--p3, respectively. These departures accompany substantial turning of the configuration-space paths. The force and work predictions differ in accuracy: on p2 at 2 fs, work differs from its directional prediction by only $-0.53\%$, while the maximum atomic residual exceeds its linear prediction by 35.1\%. The maximum-residual ordering between p2 and p3 reverses between 1 and 1.5 fs, whereas p3 continues to accumulate less positive work through 4 fs.

\section{Tungsten sampling, structural definitions and reference scope}
\label{app:tungsten-methods}
The tungsten trajectories use a lattice parameter of 3.165 \AA\ and a 0.5-fs integration timestep. The independent bulk segments have nominal 300, 3000 and 6000 K settings and are saved every 10 fs through 110 fs. The spike is saved every 25 fs through 300 fs. Times follow the archived sampling scripts and are measured separately for each independent segment. The 40 atoms within 5.5 \AA\ of the initial cell center define the core; their identities are retained throughout the trajectory. Initial positions are identical across the four segments; repeated coordinates are counted once when reporting distinct geometries. The nominal 8000-K core setting gives an actual initial regional kinetic temperature of about 10,517 K after regional center-of-mass removal. These temperatures specify the initial kinetic excitation.

Each atom's displacement is its minimum-image position difference from the segment's initial frame, with whole-cell translational drift removed. The regional RMS is $[N_R^{-1}\sum_{i\in R}|\bm u_i|^2]^{1/2}$.

The existing Quantum ESPRESSO~\cite{giannozzi2017advanced} reference labels use PBE with D3, $\Gamma$ sampling, 70/700-Ry cutoffs, 0.01-Ry Marzari--Vanderbilt smearing and $10^{-6}$-Ry electronic convergence, with a 14-valence-electron W dataset. The 432-atom inputs use 3074 bands. Reference-force RMS is the vector quantity $[N_R^{-1}\sum_{i\in R}|\bm F_i^{\rm ref}|^2]^{1/2}$; residual RMS replaces $\bm F_i^{\rm ref}$ by $\bm F_i^{\rm sur}-\bm F_i^{\rm ref}$. When the data were first saved, 27 successful label records covered 24 distinct geometries; 17 other frame attempts had failed, two were active and three had not started. The subsequent frozen-model evaluation succeeded on all 24 available geometries without adding reference labels. The force comparison therefore includes only geometries whose reference calculations completed successfully. Complete historical electronic-output files are not retained for every label; the saved force records retain their original frame and source indices.

We evaluate two locally cached models, MACE-MP-0b3-medium and MACE-MPA-0-medium, with the original committee recipe and seed 20260834. The second member retains the recipe's 1\% initialization perturbation to readout/product parameters. No W labels are used to train, calibrate or select this committee, and all effective weights are held fixed. Pretraining overlap is unknown. The four identical initial geometries use frame 0 as the predetermined primary reference; all four original labels are retained, with maximum mutual force difference 0.001911 eV/\AA. Source model and effective tensor hashes accompany the data. The force comparison uses these fixed evaluation models. Residuals include the reference's explicit D3 term, which the committee omits; the control below separates its contribution.

To measure the effect of the initialization perturbation, we evaluate both original model files without it on the shared initial lattice and the 50-fs spike configuration used in Fig.~\ref{fig:materials-tungsten}. The reference labels, atom identities and 0.2-eV/\AA\ budget are unchanged. Effective model tensors match the original files exactly, and the unchanged MP-0b3 member reproduces the archived forces component by component. At 50 fs, the original MP-0b3 and MPA-0 models have maximum residuals of 2.342 and 1.536 eV/\AA, respectively; their equal-weight mean has maximum residual 1.604 eV/\AA\ and vector RMS 0.414 eV/\AA. The corresponding mean core and matrix RMS values are 0.898 and 0.326 eV/\AA. For MPA-0 alone they are 0.688 and 0.182 eV/\AA. The perturbation changes the second member's forces by maximum 0.166 eV/\AA\ and vector RMS 0.036 eV/\AA. The large residual in the spike remains concentrated in the core when the perturbation is removed. All original-model residual maxima at the shared initial lattice are approximately 0.00199 eV/\AA.

We also compute the additive D3 contribution on this same fixed pair of geometries using the installed Quantum ESPRESSO 7.5 D3 library, with PBE zero damping, the three-body term, a $\sqrt{9000}$-bohr interaction cutoff and a 40-bohr coordination cutoff. Its initial-geometry energy reproduces the independent printed QE D3 energy within $2.1\times10^{-9}$ Ry. Along the fixed displacement direction between the two structures, central energy differences at 0.001 and 0.0005 bohr agree with the analytic gradient projection within the predeclared tolerance $10^{-6}+10^{-4}|\partial_u E_{\rm D3}|$ in Hartree/bohr. At 50 fs, the D3 force has maximum 0.06548 eV/\AA\ and vector RMS 0.01952 eV/\AA. Subtracting it from the original reference gives $\bm F_{\rm PBE}=\bm F_{\rm PBE+D3}-\bm F_{\rm D3}$. Against this electronic reference, the original MP-0b3, MPA-0 and equal-weight mean have maximum residuals of 2.306, 1.540 and 1.625 eV/\AA, respectively. The mean core and matrix residual RMS values are 0.887 and 0.319 eV/\AA. The shared initial lattice remains within budget. This decomposition quantifies the D3 contribution while retaining the other electronic-reference settings.

\paragraph*{Computational assistance.}
The authors directed the study and take responsibility for the scientific interpretation and final manuscript. Large language model tools assisted code development and debugging, data organization and analysis, checks of derivations, and manuscript drafting and language revision.

\statementheading{Data and code availability}
This manuscript is currently under peer review; access details for the data and code supporting this work will be provided in the published article.

\statementheading{Author contributions}
P. K. led the conception and design of the study, developed the theoretical framework and computational protocols, implemented the analyses, performed the principal calculations and validation, and wrote the manuscript. \mbox{V.~M.-R.} provided code and technical support for the density-functional-theory and molecular-dynamics programs. L.~Z. and \mbox{L.-D.~Z.} secured funding and provided conceptual guidance. D. W., S. B., Z. L. and Y. L. contributed to data collection and selected validation calculations.

\statementheading{Competing interests}
The authors declare no competing interests.

\bibliography{refs}
\end{document}